\documentclass[11pt]{article}
\pdfoutput=1 
\usepackage{amssymb}
\usepackage{amsmath}
\usepackage{amstext}
\usepackage{graphicx,epsfig}
\usepackage{epsfig}
\usepackage{verbatim} 
\usepackage{fancybox}
\usepackage{color}
\usepackage{ulem}
\usepackage{enumitem}
\usepackage{youngtab}
\usepackage{subfigure}
\usepackage{bbm}
\usepackage{parskip}
\usepackage[numbers,sort&compress]{natbib}
\usepackage{ytableau}
\usepackage[utf8]{inputenc}

\usepackage{tikz}
\usetikzlibrary{positioning, trees,decorations, decorations.markings, decorations.pathmorphing, arrows, shapes.geometric, snakes, patterns}
\pgfdeclaredecoration{complete sines}{initial}
{
    \state{initial}[
        width=+0pt,
        next state=upsine,
        persistent precomputation={\pgfmathsetmacro\matchinglength{
            \pgfdecoratedinputsegmentlength / int(\pgfdecoratedinputsegmentlength/\pgfdecorationsegmentlength)}
            \setlength{\pgfdecorationsegmentlength}{\matchinglength pt}
        }] {}
    \state{upsine}[width=\pgfdecorationsegmentlength,next state=downsine]{
        \pgfpathsine{\pgfpoint{0.25\pgfdecorationsegmentlength}{0.5\pgfdecorationsegmentamplitude}}
        \pgfpathcosine{\pgfpoint{0.25\pgfdecorationsegmentlength}{-0.5\pgfdecorationsegmentamplitude}}
    }
    \state{downsine}[width=\pgfdecorationsegmentlength,next state=upsine]{
        \pgfpathsine{\pgfpoint{0.25\pgfdecorationsegmentlength}{-0.5\pgfdecorationsegmentamplitude}}
        \pgfpathcosine{\pgfpoint{0.25\pgfdecorationsegmentlength}{0.5\pgfdecorationsegmentamplitude}}
}
    \state{final}{}
}

\DeclareMathOperator*{\res}{Res}
\newcommand{\Comment}[1]{{}}
\definecolor{darkblue}{rgb}{0.15,0.35,0.55}
\definecolor{reddish}{rgb}{0.65, 0.2, 0.2}
\usepackage[linktocpage=true]{hyperref}
\hypersetup{
colorlinks=true,
citecolor=darkblue,
linkcolor=reddish,
urlcolor=darkblue,
pdfauthor={},
pdftitle={},
pdfsubject={}
}

\definecolor{green3}{RGB}{44, 160, 44}

\newcommand{\be}{\begin{equation}}
\newcommand{\ee}{\end{equation}}
\newcommand{\bea}{\begin{eqnarray}}
\newcommand{\eea}{\end{eqnarray}}
\newcommand{\beas}{\begin{eqnarray*}}
\newcommand{\eeas}{\end{eqnarray*}}
\newcommand{\nn}{\nonumber}

\newcommand{\ccdot}{\! \cdot \!}

\definecolor{darkred}{rgb}{0.7,0.3,0.3}
\definecolor{darkgreen}{rgb}{0.2,0.7,0.3}
\definecolor{lightgreen}{rgb}{.816,.94,.753}
\definecolor{greyish}{rgb}{.8,.8,.8}
\definecolor{darkblue2}{rgb}{0.3,0.4,0.9}
\usepackage{pifont}
\usepackage{xcolor,colortbl}

\def\({\left(}
\def\){\right)}

\newcommand{\rd}{{\rm d}}

\def\gsim{ \lower .75ex \hbox{$\sim$} \llap{\raise .27ex \hbox{$>$}} }
\def\lsim{ \lower .75ex \hbox{$\sim$} \llap{\raise .27ex \hbox{$<$}} }

\def\xyma{\xymatrix@M.7em}
\def\xymas{\xymatrix@M.1em}

\title{}
\author{}

\numberwithin{equation}{section}

\begin{document}
\tikzset{
    photon/.style={decorate, decoration={snake}, draw=magenta},
    graviton/.style={decorate, decoration={snake}, draw=black},
 sgal/.style={decorate , dashed, draw=black},
 scalar/.style={decorate , draw=black},
  mgraviton/.style={decorate, draw=black,
    decoration={coil,amplitude=4.5pt, segment length=7pt}}
    electron/.style={draw=blue, postaction={decorate},
        decoration={markings,mark=at position .55 with {\arrow[draw=blue]{>}}}},
    gluon/.style={decorate, draw=magenta,
        decoration={coil,amplitude=4pt, segment length=5pt}} 
}
\renewcommand{\thefootnote}{\fnsymbol{footnote}}
~

\begin{center}
{\huge \bf Matter Couplings and Equivalence Principles\\
\vskip4pt
 for Soft Scalars}
\end{center} 

\vspace{1truecm}
\thispagestyle{empty}
\centerline{\Large James Bonifacio,${}^{\rm a,}$\footnote{\href{mailto:james.bonifacio@case.edu}{\texttt{james.bonifacio@case.edu}}} Kurt Hinterbichler,${}^{\rm a,}$\footnote{\href{mailto:kurt.hinterbichler@case.edu} {\texttt{kurt.hinterbichler@case.edu}}} Laura A. Johnson,${}^{\rm a,}$\footnote{\href{mailto:lxj154@case.edu} {\texttt{lxj154@case.edu}}} }

\vskip 4pt
\centerline{\Large
Austin Joyce,${}^{\rm b,}$\footnote{\href{mailto:austin.joyce@columbia.edu}{\texttt{austin.joyce@columbia.edu}}} and Rachel A. Rosen${}^{\rm b,}$\footnote{\href{mailto:rar2172@columbia.edu}{\texttt{rar2172@columbia.edu}}}}
\vspace{.5cm}

\centerline{{\it ${}^{\rm a}$CERCA, Department of Physics,}}
 \centerline{{\it Case Western Reserve University, 10900 Euclid Ave, Cleveland, OH 44106}} 
 \vspace{.25cm}
 
 \centerline{{\it ${}^{\rm b}$Center for Theoretical Physics, Department of Physics,}}
 \centerline{{\it Columbia University, New York, NY 10027}} 

 \vspace{1cm}
\begin{abstract}
\noindent
Scalar effective field theories with enhanced soft limits behave in many ways like gauge theories and gravity. 
In particular, symmetries fix the structure of interactions and the tree-level $S$-matrix in both types of theories. 
We explore how this analogy persists in the presence of matter by considering theories with additional fields coupled to the Dirac--Born--Infeld (DBI) scalar or the special galileon in a way that is consistent with their symmetries.
Using purely on-shell arguments, we show that these theories obey analogues of the $S$-matrix equivalence principle whereby all matter fields must couple to the DBI scalar or the special galileon through a particular quartic vertex with a universal coupling. 
These equivalence principles imply the universality of the leading double soft theorems in these theories, which are scalar analogues of Weinberg's gravitational soft theorem, and can be used to rule out interactions with massless higher-spin fields when combined with analogues of the generalized Weinberg--Witten theorem. 
We verify in several examples that amplitudes with external matter fields nontrivially exhibit enhanced single soft limits and we show that such amplitudes can be constructed using soft recursion relations when they have sufficiently many external DBI or special galileon legs, including amplitudes with massive higher-spin fields. As part of our analysis we construct a recently conjectured special galileon-vector effective field theory.
\end{abstract}

\newpage

\setcounter{tocdepth}{2}
\tableofcontents
\renewcommand*{\thefootnote}{\arabic{footnote}}
\setcounter{footnote}{0}

\newpage
\section{Introduction}

Aspects of gravitational physics remain mysterious even today, a century after the discovery of Einstein gravity. 
At extremely high energies---where the dynamics becomes necessarily quantum---our understanding of gravity is certainly incomplete. There is also the tantalizing possibility that there is more to learn about gravity in the infrared, and that the observed cosmic acceleration is the first hint of some new physics.  Moreover, a thorough understanding of black hole information remains elusive.
In light of this, it is worthwhile to broadly explore avenues toward decoding the gravitational sector.

In this paper, we explore an analogy between gravity and scalar field theories that obey certain soft theorems.
These {\it soft scalars} have scattering amplitudes that vanish like a power law when one of the external momenta is taken to zero, generalizing the Adler zero. It has recently been understood that this behavior completely fixes the $S$-matrix of the scalar, under mild assumptions~\cite{Cheung:2014dqa, Cheung:2015ota, Luo:2015tat,Cheung:2016drk,Padilla:2016mno,Elvang:2018dco, Low:2019ynd, Rodina:2018pcb}. This can also be understood in terms of the presence of symmetries which constrain scalar self-interactions~\cite{Nicolis:2008in,deRham:2010eu,Hinterbichler:2014cwa,Griffin:2014bta,Hinterbichler:2015pqa,Novotny:2016jkh,Bogers:2018zeg,Roest:2019oiw}. 
This rigid structure is reminiscent of the way in which gauge invariance fixes the self-interactions of the graviton, along with its leading interactions with matter fields. We will see that this analogy goes beyond the superficial level, and that many of the interesting structures discovered within gravity have echoes in soft scalar theories, as summarized in Table~\ref{tab:dictionary}.

The application of scattering amplitude techniques to gravity brings certain features to the fore that would 
otherwise be less obvious. Most famously, Weinberg used on-shell gauge invariance to derive the universality of gravity's coupling to matter---the equivalence principle---from $S$-matrix factorization in the soft limit~\cite{Weinberg:1964ew}, and it has recently been understood that there are a number of similar subleading soft theorems~\cite{Cachazo:2014fwa,Schwab:2014xua,Campiglia:2014yka}. Additionally, the fact that graviton self-interactions are completely fixed at lowest derivative order by gauge invariance manifests from the $S$-matrix perspective as the on-shell recursive constructibility of scattering amplitudes~\cite{Britto:2004ap,Britto:2005fq,Cachazo:2005ca,Benincasa:2007qj,ArkaniHamed:2008yf}. We will see that each of these features has a precise soft scalar analogue.

In order to explore the analogous behaviors in soft scalar theories, we examine two concrete examples: 
the Dirac--Born--Infeld (DBI) theory and the special galileon. Each of these theories has a nonlinear shift symmetry that both fixes their structure~\cite{deRham:2010eu,Hinterbichler:2015pqa} and  protects the soft behavior of their amplitudes. To fully explore the analogy with gravity, it will be important to couple these theories to additional matter fields. From this perspective, the distinguishing feature of these theories is the existence of a {\it shift-covariant} effective metric to which matter couples (see Table~\ref{tab:dictionary} for the explicit expressions). Under a shift symmetry transformation, this metric transforms by the Lie derivative along some particular vector field. The fact that the effective metric is covariant (rather than invariant) tightly constrains the possible interactions between these Goldstone fields and matter, since the matter fields must also transform under the shift symmetries.
We explore various aspects of the coupled scalar-matter systems. In particular, we verify that coupling to matter preserves the single-soft behavior for external scalar legs. As a byproduct, we are able to explicitly construct a coupled special galileon-vector theory which was recently conjectured to exist~\cite{Elvang:2018dco,Roest:2019oiw}.

Each of the interesting features of gravitational scattering amplitudes has an analogue in the context of the DBI and special galileon theories. For example, we will see that because of the rigid structure imposed by symmetry, these scalar field theories have universal leading-order couplings to matter, which implies a precise analogue of the equivalence principle.  The derivation of this statement from the scattering viewpoint parallels the derivation of Weinberg's $S$-matrix equivalence principle and soft graviton theorem, with double-soft factorization for the scalar playing the same role as single-soft factorization for the graviton. From this factorized statement, the additional information that is analogous to the on-shell Ward identity is the demand that the universal soft factor satisfies the appropriate single-soft theorem. Similar to the subleading single soft theorems enjoyed by gravity \cite{Cachazo:2014fwa,Schwab:2014xua,Campiglia:2014yka}, the DBI and special galileon theories also satisfy subleading double soft theorems~\cite{Cachazo:2015ksa,Li:2017fsb,Guerrieri:2017ujb}. We verify that these subleading double soft theorems continue to hold in the presence of matter couplings, while the sub-subleading theorems are not universal.

The on-shell constructibility of tree-level gravity amplitudes is an important feature of the theory. Recently it has been understood that soft scalar field theories can similarly be recursively constructed---a program known as the {\it soft bootstrap}~\cite{Cheung:2016drk,Padilla:2016mno,Elvang:2018dco, Low:2019ynd}. Here we explore the extent to which DBI and the special galileon can continue to be constructed recursively when coupled to matter fields. As an example, we find that it is possible to bootstrap all amplitudes with a sufficient number of DBI or special galileon legs in theories of free matter fields minimally coupled to the DBI scalar or special galileon.

\bgroup
\def\arraystretch{1.25}%
\begin{table}[!t]
  \centering
{\scriptsize
 \begin{tabular}{  l | c | c | c }
 ~ & Einstein gravity & DBI& Special galileon \\ \hline
 Lagrangian:& \raisebox{-0.2ex}{$\sqrt{-g} R$} & \raisebox{-0.3ex}{$-\sqrt{1+(\partial\phi)^2}$} & $\mathcal{L}_{\rm sgal}$ \\
Structural symmetry:&   Diffeomorphisms & Higher-dimensional boosts & Quadratic shift symmetry\\ 
Metric: & $ g_{\mu\nu}$ & $\eta_{\mu\nu}+\partial_\mu\phi\partial_\nu\phi$ & $\eta_{\mu\nu}-\partial_\mu\partial_\rho\phi\partial^\rho\partial_\nu\phi$ \\
On-shell constraint:& On-shell Ward identity & Vanishing single soft theorem & Vanishing single soft theorem \\
Soft factorization:&    Single soft theorem &Double soft theorem& Double soft theorem \\
 Equivalence principle:& $S$-matrix equiv. principle &DBI equiv. principle& Special galileon equiv. principle \\
 Constructibility:& BCFW recursion &Soft recursion& Soft recursion\\
 Double copy:& YM $\otimes$ YM &YM $\otimes$ NLSM & NLSM $\otimes$ NLSM 
  \end{tabular}
  }
  \caption{\small Summary of the analogy between Einstein gravity, DBI theory, and the special galileon. Many of the defining features of gravity from the scattering perspective have precise analogues within the scalar theories. The Lagrangian of the special galileon, $\mathcal{L}_{\rm sgal}$, is defined in Eq.~\eqref{eq:sgal}. Note that compared to the main text we have set $\alpha=\Lambda=1$. The acronym YM stands for Yang--Mills, while NLSM stands for nonlinear sigma model.
  } \label{tab:dictionary}
\end{table}
\egroup

Given the close analogy between the soft scalars and gravity, we also indulge in some speculation about how gravity would work in a world where there was no graviton, but instead one of the scalars mediated the gravitational force. In this hypothetical world there would be some welcome features; for example, there is in a precise sense no cosmological constant (CC) problem.  Unfortunately, there are also some less realistic and unwelcome features; for example, the Newtonian gravitational potential would fall off like $\sim r^{-11}$ for the special galileon and like $\sim r^{-7}$ for the DBI scalar.

It is worth noting that many analogies and direct correspondences between gravity and various scalar field theories have been considered before. See, for example, Refs.~\cite{Sundrum:2003yt,Coradeschi:2013gda, Cachazo:2014xea,  Cachazo:2015ksa, Cheung:2015ota, Klein:2015iud, Agrawal:2016ubh, Cheung:2016prv, Cheung:2016drk, Cheung:2017ems, Cheung:2017yef, Zhou:2018wvn, Bollmann:2018edb}. Our focus is to emphasize the universal coupling of soft scalar field theories to additional matter fields, particularly from the $S$-matrix point of view.

The broad outline of the paper is the following: We begin by describing in Section~\ref{sec:coupling} the construction of matter couplings consistent with the DBI and special galileon symmetries. In Section~\ref{sec:singlesoft} we verify that these couplings do not spoil the single soft behavior of the Goldstone theories. We then derive a version of the equivalence principle for DBI and the special galileon in Section~\ref{sec:equiv}. 
An output of this derivation is the universality of the leading double soft theorems previously derived for pure scalar theories. We also show that these equivalence principles are incompatible with massless higher-spin particles by proving an analogue of the generalized Weinberg--Witten theorem. In Section~\ref{sec:recursion} we consider the recursive construction of the $S$-matrix for theories involving additional matter fields interacting with the DBI scalar or the special galileon, including massive fields with arbitrary integer spin. We consider some phenomenological aspects of these scalar gravitational theories in Section~\ref{eq:fakegravity}, although they are not realistic. We collect some technical results in the Appendices.

\vspace{.15cm}
\noindent
{\bf Conventions:}
We work in $D$ spacetime dimensions with $D\geq 3$ and use the mostly-plus metric signature convention. In scattering amplitudes all momenta are defined to be incoming and we always replace symmetric traceless polarization tensors with products of null vectors, $\epsilon_i^{\mu_1 \dots \mu_s} \mapsto \epsilon_i^{\mu_1} \dots \epsilon_i^{\mu_s}$ where $\epsilon_i \cdot \epsilon_i =0$.  We denote dot products between momenta by $p_{ab} \equiv p_a\cdot p_b$.

\section{Coupling to matter}
\label{sec:coupling}
To fully explore the analogy between gravity and certain scalar theories with enhanced soft limits, it is essential to couple the scalar theories to matter fields while retaining their shift symmetries. This is the analogue of coupling gravity to matter in a diffeomorphism-invariant way. In this Section, we review how---as for gravity---there is a metric built from the relevant fields that transforms covariantly under the shift symmetries. This metric can thus be used to couple to matter in a way that preserves the symmetries, provided that we transform the matter fields in an appropriate way.

\subsection{DBI theory}
We first consider the DBI scalar field theory \cite{Born:1934gh,Dirac:1962iy}. This theory is described by the action\footnote{There are possible higher-derivative terms compatible with the symmetries \cite{deRham:2010eu}, but we focus on the leading-order interactions.}
\be
S_{\rm DBI} = -\frac{\Lambda^{D}}{\alpha}\int\rd^Dx \sqrt{1+\frac{\alpha}{\Lambda^D} (\partial\phi)^2}\,,
\label{eq:DBIlag}
\ee
where we have introduced the energy scale $\Lambda$, which together with the dimensionless parameter $\alpha$ sets the scale of strong coupling. The strong coupling scale is the only free parameter, but we have introduced $\alpha$ separately as it will sometimes be useful to count factors of $\alpha$ and because its sign can be important. 
Expanding out the first few terms gives
\be \label{eq:dbi-expanded}
S_{\rm DBI} =\int\rd^Dx \left(  -\frac{1}{2}(\partial \phi)^2+ \frac{\alpha }{8 \Lambda^D} (\partial \phi)^4- \frac{\alpha^2}{16\Lambda^{2D}} (\partial \phi)^6 + \frac{5\alpha^3}{128\Lambda^{3D}} (\partial \phi)^8+ \dots \right)\,. 
\ee
The action~\eqref{eq:DBIlag} is invariant under two types of nonlinearly realized symmetries: one is a shift by a constant $c$,
\be \delta\phi = c,\label{shiftsyme}\ee
and the other acts as
\be
\delta\phi = b_\mu\left(x^\mu+\frac{\alpha}{\Lambda^D}\phi\partial^\mu\phi\right),
\label{eq:DBIsymms}
\ee
where $b_{\mu}$ is a constant vector. The action also has an obvious $\mathbb{Z}_2$ symmetry under $\phi \mapsto-\phi$.

The DBI action has an interpretation as the world-volume action of a $D$-dimensional brane embedded in ${\mathbb R}^{D,1}$, where the nonlinearly realized symmetries~\eqref{eq:DBIsymms} are the higher-dimensional Lorentz transformations and the shift \eqref{shiftsyme} is the higher-dimensional translation, all of which are spontaneously broken by the presence of the brane. The ambient Minkowski metric can be pulled back to the brane, where it is given by~\cite{deRham:2010eu,Goon:2011qf}
\be
\tilde g_{\mu\nu} = \eta_{\mu\nu}+\frac{\alpha}{\Lambda^D}\partial_\mu\phi\partial_\nu\phi,
\label{eq:dbimetric}
\ee
and the DBI action \eqref{eq:DBIlag} can be written as the square root determinant of this induced metric.

This geometric interpretation provides a natural way to couple the DBI scalar field $\phi$ to additional matter fields.
The induced metric~\eqref{eq:dbimetric} is strictly invariant under the shift symmetry $\phi \mapsto \phi+c$, while under the boost-like symmetry~\eqref{eq:DBIsymms} it transforms by the Lie derivative
along the vector field $v^\mu = \alpha b^\mu\phi/\Lambda^D$~\cite{deRham:2010eu,Goon:2011qf},\footnote{Recall that the Lie derivative of any metric along a vector field $v^\mu$ can be written as
\be
{\cal L}_v g_{\mu\nu} = v^\alpha\partial_\alpha g_{\mu\nu}+ g_{\alpha\nu}\partial_\mu v^\alpha+g_{\alpha\mu}\partial_\nu v^\alpha.
\label{eq:liederiv}
\ee
}
\be \delta \tilde g_{\mu\nu}={\cal L}_{v}\tilde g_{\mu\nu}\,,\quad{\rm with}\quad v^\mu=\frac{\alpha}{\Lambda^D}b^\mu\phi\, .\ee
If we couple to additional matter fields in a diffeomorphism-invariant way using this metric, and also transform the matter fields by the Lie derivative along the direction $v^\mu$ as part of the action of the symmetry, then the theory will be invariant under the transformation.\footnote{The induced transformation properties of the matter fields can be understood in two equivalent ways. From the brane perspective, the higher-dimensional boost symmetries take us out of static gauge and require a compensating world-volume reparametrization to restore the gauge. The brane matter fields transform under this coordinate change by the Lie derivative along $v^\mu$ \cite{Gliozzi:2011hj}. Alternatively, this transformation of matter fields can be understood from the coset construction~\cite{Goon:2012dy}.}

As a simple example, we can consider coupling the DBI scalar to an additional scalar field $\chi$ with mass $m_\chi$ as
\be 
\label{eq:dbiscalar}
S_{\chi} =\int\rd^Dx\sqrt{-\tilde g} \left(- \frac{1}{2} \tilde g^{\mu \nu} \partial_{\mu} \chi \partial_{\nu} \chi -\frac{m^2_{\chi}}{2} \chi^2 \right).
\ee
This action is invariant under the DBI symmetries, provided the matter fields transform as
\be
\delta\chi ={\cal L}_v\chi= \frac{\alpha}{\Lambda^D} \phi\, b^\mu\partial_\mu\chi
\ee
under the boost symmetry~\eqref{eq:DBIsymms} and do not transform under the shift symmetry. Note that the determinant and inverse metric involve arbitrarily many even powers of $\phi$, so the action~\eqref{eq:dbiscalar} involves an infinite number of interactions between the matter field and the DBI scalar.

Couplings to other forms of matter can be engineered essentially by following the minimal-coupling prescription for gravity,  using the metric $\tilde g_{\mu\nu}$. We give additional examples below when we consider the special galileon.

\subsection{Special galileon}
Our other scalar theory of interest is the special galileon \cite{Cachazo:2014xea, Cheung:2014dqa, Hinterbichler:2015pqa}. The special galileon is a sum of all the galileon terms with even numbers of fields in $D$ dimensions, with fixed relative coefficients:\footnote{The precise admixture of galileon terms can be changed by so-called galileon duality field redefinitions~\cite{deRham:2013hsa,Kampf:2014rka}, so the more precise statement is that the special galileon admits a duality frame in which the action takes the form in Eq.~\eqref{eq:sgal}.}
\be 
\label{eq:sgal}
S_{\rm sgal} =-\frac{1}{2}\int\rd^Dx \sum_{n=1}^{ \left \lfloor \frac{D+1}{2} \right \rfloor}  \frac{\alpha^{n-1}}{(2n-1)!\Lambda^{(D+2)(n-1)}} \left( \partial \phi \right)^2 \mathcal{L}^{\rm TD}_{2n-2},
\ee
where as with DBI we have introduced an energy scale $\Lambda$ and a dimensionless parameter $\alpha$ that together set the scale of strong coupling. In Eq.~\eqref{eq:sgal}  the total derivative combinations $\mathcal{L}^{\rm TD}_{n}$ are defined by
\be
\mathcal{L}^{\rm TD}_{n} \equiv \sum_p (-1)^p \eta^{\mu_1 p({\nu_1})} \cdots \eta^{\mu_n p({\nu_n})} \Phi_{\mu_1 \nu_1} \cdots \Phi_{\mu_n \nu_n},
\ee
with $\Phi_{\mu \nu} \equiv \partial_{\mu} \partial_{\nu} \phi$, and the sum runs over all permutations of the $\nu$ indices with $(-1)^p$ the sign of the permutation.
As an explicit example, the theory in $D=4$ takes the form
\be
S = \int\rd^4x \left(-\frac{1}{2}(\partial\phi)^2 - \frac{\alpha}{12\Lambda^6}(\partial\phi)^2\Big[(\square\phi)^2 - (\partial_\mu\partial_\nu\phi)^2\Big]\right).
\label{eq:sgal4d}
\ee
In addition to the familiar shift symmetries enjoyed by the individual galileon terms \cite{Nicolis:2008in},
\be \label{eq:galshift}
\delta\phi =c+b_{\mu}x^\mu,
\ee
the structure of the action is fixed by invariance under the higher-order shift symmetry~\cite{Hinterbichler:2015pqa}
\be
\delta\phi = s_{\mu\nu} \left(x^\mu x^\nu -\frac{\alpha}{\Lambda^{D+2}}\partial^\mu\phi\partial^\nu\phi\right),
\label{eq:sgalshift}
\ee
where $s_{\mu\nu}$ is a traceless symmetric constant tensor. Lastly, like the DBI theory, the special galileon also has a $\mathbb{Z}_2$ symmetry $\phi\mapsto -\phi$.

As in the DBI theory, we can construct an effective metric from the special galileon field
\be
\bar g_{\mu\nu} = \eta_{\mu\nu}-\frac{\alpha}{\Lambda^{D+2}}\partial_\mu\partial_\alpha\phi\partial^\alpha\partial_\nu\phi,
\label{eq:sgalmetric}
\ee
which transforms covariantly under the extended shift symmetry~\eqref{eq:sgalshift}; under a shift it transforms by the Lie derivative~\eqref{eq:liederiv} along the following vector field:
\be
\delta \bar g_{\mu\nu}={\cal L}_v\bar g_{\mu\nu}\, ,\ \ \ v^\mu = -\frac{2\alpha}{\Lambda^{D+2}} s^{\mu\nu}\partial_\nu\phi\,. \label{vdefine}
\ee
This effective metric can be understood as arising either from a geometric embedding~\cite{Novotny:2016jkh}, or from the coset construction~\cite{Bogers:2018zeg,Garcia-Saenz:2019yok,Carrillo-Gonzalez:2019aao}. Note that---unlike the DBI case---the special galileon itself~\eqref{eq:sgal} {\it cannot} be written in terms of this effective metric. Rather, it is a Wess--Zumino term for the relevant symmetries~\cite{Hinterbichler:2015pqa,Bogers:2018zeg,Garcia-Saenz:2019yok}.

The covariance of the metric~\eqref{eq:sgalmetric} suggests a simple way to couple the special galileon to matter fields. As in the DBI case, we let the matter fields transform under the special galileon symmetry as the Lie derivative of the vector field $v^\mu$, defined in Eq.~\eqref{vdefine}, and then couple them by forming diffeomorphism invariants from the metric~\eqref{eq:sgalmetric}, just as we couple matter to the graviton in general relativity (GR). A few explicit examples of this general procedure are as follows:
\begin{itemize}
\item {\bf Scalar:} We can add to $S_{\rm sgal}$ a minimally coupled scalar field $\chi$ with mass $m_\chi$ with the action\footnote{The inverse metric $\bar g^{\mu\nu}$ that appears in the action is given by~\cite{Novotny:2016jkh}
\be
\bar g^{\mu\nu} = \eta^{\mu\nu} + \sum_{n=1}^\infty \frac{\alpha^n}{\Lambda^{(D+2)n}} \Phi^{\mu\alpha} (\Phi^{2n-2})_{\alpha\beta} \Phi^{\beta\nu} ,
\ee
where $\Phi_{\mu \nu} \equiv \partial_{\mu} \partial_{\nu} \phi$ and $(\Phi^{n})_{\mu\nu} = \Phi_{\mu\alpha_1}\Phi^{\alpha_1}_{\alpha_2}\cdots\Phi^{\alpha_{n-1}}_{\nu}$. In this expression the $\Phi$ indices are raised and lowered using the flat metric.
}
\be \label{eq:sgalscalar}
S_{\chi} =- \frac{1}{2}\int\rd^Dx\sqrt{-\bar g}\,\Big( \bar g^{\mu \nu} \partial_{\mu} \chi \partial_{\nu} \chi + m^2_{\chi} \chi^2 \Big),
\ee
where the metric $\bar g_{\mu\nu}$ is given by~\eqref{eq:sgalmetric} and the scalar field $\chi$ transforms under the special galileon symmetry as the scalar Lie derivative along the vector field in Eq.~\eqref{vdefine},
\be
\delta \chi  = \mathcal{L}_v \chi  =  -\frac{2\alpha}{\Lambda^{D+2}} s^{\mu \nu} \partial_{\mu} \phi \partial_{\nu} \chi\,.
\ee
As for the DBI theory, the action~\eqref{eq:sgalscalar} involves an infinite number of interactions between the matter field and the special galileon due to the determinant and inverse metric.

\item {\bf Vector:} We can add to $\mathcal{S}_{\rm sgal}$ a minimally coupled spin-1 particle $A_{\mu}$ with mass $m_A$,
\be \label{eq:sgalvector}
S_{A} = \int\rd^Dx\sqrt{-\bar g} \left( - \frac{1}{4}F_{\mu \nu} F^{\mu\nu} -\frac{m^2_A}{2}  A_{\mu} A^{\mu} \right),
\ee
where $F_{\mu \nu } \equiv \partial_{\mu} A_{\nu} - \partial_{\nu} A_{\mu}$ and indices are raised and lowered with $\bar g_{\mu\nu}$. The vector field transforms under the special galileon symmetry as
\be
\delta A_\mu  = \mathcal{L}_v A_{\mu} = -\frac{2\alpha}{\Lambda^{D+2}} s^{\alpha\beta}\left(\partial_\beta\phi\partial_\alpha A_\mu+\partial_\beta\partial_\mu\phi A_\alpha \right),
\ee
which is the same transformation found by different arguments in Ref.~\cite{Roest:2019oiw}. We show later that in the massless case, $m_A = 0$, this theory is the conjectured vector--special galileon theory whose $S$-matrix was partially constructed in Ref.~\cite{Elvang:2018dco} through soft bootstrap arguments. 
Again, note that the action~\eqref{eq:sgalvector} has interactions involving arbitrarily many even powers of $\phi$.
\item {\bf Spin-2:} We can add a massive spin-2 particle $H_{\mu \nu}$ with the following addition to the action\footnote{Note that we have not included non-minimal couplings between the massive spin-2 particle and the background curvature associated to the metric $\bar g_{\mu\nu}$. Such couplings are typically required to ensure that the massive spin-2 particle propagates the correct number of degrees of freedom \cite{Bernard:2014bfa,Bernard:2015mkk,Bernard:2015uic,Bernard:2017tcg}. Since we are considering everything in perturbation theory, these considerations will not be important to us---as a consequence this coupling will not be ghost free, but this does not affect the perturbative computation of the $S$-matrix. 
See Appendix~\ref{sec:dof} for  more discussion.} 
\begin{align}
S_{H} = \int\rd^Dx \sqrt{-\bar g} \bigg(-\frac{1}{2} \bar\nabla_{\lambda} H_{\mu\nu} \bar\nabla^{\lambda} H^{\mu\nu} +
\bar\nabla_{\mu} H_{\nu \lambda} \bar\nabla^\nu H^{\mu\lambda}& -\bar\nabla_\mu H\bar\nabla_\nu H^{\mu\nu}+\frac{1}{2}\bar\nabla_\mu H\bar\nabla^\mu H \nn \\
&-\frac{m_H^2}{2}  \left(H_{\mu\nu}H^{\mu\nu} -H^2 \right) \bigg),
\end{align}
where $\bar\nabla$ is the covariant derivative of $\bar g_{\mu \nu}$ and indices are again raised and lowered with $\bar g_{\mu\nu}$. The massive spin-2 field transforms as
\be
\delta H_{\mu \nu} ={\cal L}_vH_{\mu\nu}= -\frac{2\alpha}{\Lambda^{D+2}} s^{\sigma\lambda}\partial_\lambda\phi\partial_\sigma H_{\mu\nu}- H_{\sigma\nu}\partial_\mu \left(\frac{2\alpha}{\Lambda^{D+2}} s^{\sigma\lambda}\partial_\lambda\phi\right)-H_{\sigma\mu}\partial_\nu\left(\frac{2\alpha}{\Lambda^{D+2}} s^{\sigma\lambda}\partial_\lambda\phi\right)\,.
\ee
\end{itemize}

In addition to the minimally coupled free fields described above, it is also possible to add
matter self-interactions and non-minimal terms containing the Riemann curvature of $\bar g_{\mu \nu}$, e.g.,
\begin{align}
\Delta S_{\chi} =& \int\rd^Dx  \sqrt{-\bar g}\left(\frac{\lambda_n}{n!} \chi^n + \lambda (\partial \chi)^2 \Box \chi+ \rho  \chi^2 R(\bar{g})\right) \,. \label{eq:scalarselfint}
\end{align}
In general, we can use the metrics $\tilde{g}_{\mu \nu}$ and $\bar{g}_{\mu \nu}$ to couple the DBI scalar and the special galileon to anything to which gravity can couple. Since diffeomorphism invariance is incompatible with higher-spin gauge symmetry in flat space, we expect that these scalars cannot couple to massless spin-$s$ particles with $s \geq 2$ while preserving all the various shift symmetries, as we discuss more below. However, there is no symmetry obstruction to coupling DBI or the special galileon to massive particles of any spin, including massive gravity~\cite{deRham:2010kj}.
Note also that we cannot couple two different special galileon fields~\cite{Elvang:2018dco, Roest:2019oiw}, just as we cannot couple two gravitons~\cite{Boulanger:2000rq}. There do exist multi-DBI theories \cite{Hinterbichler:2010xn}, but in this case there is still a unique covariant metric. There also exists a theory involving the special galileon interacting with NLSM and biadjoint scalars that controls the single soft limits of special galileon amplitudes \cite{Cachazo:2016njl}.

\section{Single soft limits}
\label{sec:singlesoft}

A crucial ingredient in the construction of gauge theory amplitudes is the requirement that on-shell amplitudes are gauge invariant. Along with Lorentz invariance, this fixes a large measure of the structure of the theory~\cite{Arkani-Hamed:2016rak}. Most strikingly, these combined principles completely fix the structure of on-shell three-particle amplitudes, which then form the seeds from which the theory can be recursively generated. In the context of the shift-symmetric scalars we are studying, an analogous role is played by the single soft behavior of the theories, since demanding a particular generalized Adler zero in the single soft limit completely fixes the lowest-order interactions. We will later review how the full $S$-matrix can be grown from this seed.

In order to fix notation, we first review the systematics of taking soft limits of scattering amplitudes. Consider an $N$-point scattering amplitude, ${\cal A}_N $. We rescale one of the external momenta, $p_a\mapsto \tau p_a$, and then take the limit $\tau \to 0$. The scattering amplitude in this soft limit takes the schematic form
\be
{\cal A}_N \sim \tau^\sigma\left(\,\cdots\right)+\cdots.
\ee
Here the parameter $\sigma$ characterizes the softness of the amplitude, with larger positive values indicating that the amplitude goes to zero more rapidly as we scale the external momentum to zero. To make this unambiguous we use momentum conservation to eliminate other momenta in favor of $p_a$, thus maximizing $\sigma$.

Given this definition of the soft limit, we can ask for a classification of theories based on the power of $\sigma$ they display \cite{Cheung:2014dqa,Cheung:2016drk,Padilla:2016mno,Elvang:2018dco,Low:2019ynd, Rodina:2018pcb}---one may think of this as a generalization of the Adler zero condition. To get nontrivial results some restrictions must be placed on the theories, otherwise we can get any soft behavior by including many derivatives. The natural constraint is to limit the number of derivatives per field that appear in the action. With this restriction, theories that have softer-than-expected behavior have enhanced symmetries, which enforce cancellations between Feynman diagrams with different topologies in the soft limit. This makes the study of soft limits nicely complementary to the parallel effort to classify theories with extended shift symmetries~\cite{Hinterbichler:2014cwa,Griffin:2014bta,Hinterbichler:2015pqa,Novotny:2016jkh,Bogers:2018zeg,Roest:2019oiw}.

In this Section, we study the single soft limits of scattering amplitudes of DBI and the special galileon. Our motivation is two-fold. First, we want to review how the Adler zero condition constrains the structure of theses theories. Secondly,
we want to verify explicitly that the matter couplings introduced in Section~\ref{sec:coupling} preserve the enhanced Adler zero that these theories have in isolation. In particular, since the matter interactions we consider are constructed to preserve the DBI or special galileon symmetry, the resulting amplitudes should have enhanced soft behavior when the DBI or special galileon legs are taken soft. This only holds for all interactions if we impose that they preserve the $\mathbb{Z}_2$ symmetry, since otherwise there can be higher-derivative cubic interactions that spoil the vanishing single soft behavior~\cite{Cheung:2016drk,Carrillo-Gonzalez:2019aao,Kampf:2019mcd}.

\subsection{DBI theory}
We start by briefly considering soft limits of amplitudes in the pure DBI theory. The quartic interaction in Eq.~\eqref{eq:dbi-expanded} gives the amplitude
\be
\mathcal{A}( 1_\phi, 2_\phi, 3_\phi, 4_\phi) = \frac{\alpha}{\Lambda^D}\left(p_{12}^2+p_{13}^2+p_{14}^2 \right).
\ee
This amplitude trivially has $\sigma = 2$ soft behavior for each leg, which can be seen after using conservation of momentum. There are many equivalent ways to write on-shell amplitudes using momentum conservation, but it will be convenient for our later considerations to write this amplitude in the following form:
\be \label{eq:DBIquartic}
\mathcal{A}( 1_\phi, 2_\phi, 3_\phi, 4_\phi) = -\frac{2\alpha}{\Lambda^D}\left(p_{13} p_{23} -p_{12}^2\right),
\ee
which has the terms ordered by their total degree in $p_1$ and $p_2$ while also being manifestly symmetric under $p_1 \leftrightarrow p_2$. From now on we will always write quartic amplitudes and vertices this way, since this manifests the relative importance of each term in the double soft expansions that we consider later.

The first nontrivial features occur at six points. At this order there are two distinct contributions to scattering: there is an exchange contribution from the $(\partial\phi)^4$ vertex and there is a contact contribution from the $(\partial\phi)^6$ vertex, as depicted in Figure~\ref{fig:6ptsgal}. Individually, these contributions have $\sigma = 1$, but the leading-order pieces in the soft limit cancel against each other so that the full amplitude has a $\sigma = 2$ soft limit.
We can similarly calculate the eight-point DBI amplitude to see that there are nontrivial cancellations between diagrams to achieve the $\sigma =2$ soft behavior of the total amplitude.

\subsection{Special galileon}

Now we consider pure special galileon amplitudes. 
The first interaction occurs at quartic order and the corresponding on-shell four-point amplitude is
\be \label{eq:sgal4pt}
\mathcal{A}( 1_\phi, 2_\phi, 3_\phi, 4_\phi) = -\frac{2\alpha}{\Lambda^{D+2}}  p_{12}\, p_{13}\, p_{23}\,.
\ee
This amplitude trivially has $\sigma =3$ soft behavior for each leg. 
\begin{figure}[!t]
\begin{center}
\resizebox{7.25cm}{!}{
\begin{tikzpicture}[thick, node distance=1.55cm and 1.55cm,scale=.2,line width=1.3 pt]

\coordinate (vertex1);
\coordinate[above= of vertex1] (e1);
\coordinate[left=of vertex1] (e2);
\coordinate[below= of vertex1] (e3);
\coordinate[right=1.5cm of vertex1] (vertex2);
\coordinate[above= of vertex2] (e4);
\coordinate[right =of vertex2] (e5);
\coordinate[below= of vertex2] (e6);
\coordinate[right =.75cm of vertex1] (midpoint1);

\draw[sgal] (e1) -- (vertex1);
\draw[sgal] (e2) -- (vertex1);
\draw[sgal] (e3) -- (vertex1);
\draw[sgal] (vertex1) -- (vertex2);
\draw[sgal] (e4) -- (vertex2);
\draw[sgal] (e5) -- (vertex2);
\draw[sgal] (e6) -- (vertex2);

\coordinate[right=7cm of vertex1](vertex3);
\coordinate[above left=2cm of vertex3] (f1);
\coordinate[above=of vertex3] (f2);
\coordinate[above right=2cm of vertex3] (f3);
\coordinate[below right=2cm of vertex3] (f4);
\coordinate[below =of vertex3] (f5);
\coordinate[below left=2cm of vertex3] (f6);

\draw[sgal] (f1) -- (vertex3);
\draw[sgal] (f2) -- (vertex3);
\draw[sgal] (f3) -- (vertex3);
\draw[sgal] (f4) -- (vertex3);
\draw[sgal] (f5) -- (vertex3);
\draw[sgal] (f6) -- (vertex3);

\end{tikzpicture}
}
\end{center}
\caption{\small Six-point diagrams for pure DBI/special galileon amplitudes. For DBI, the exchange and contact contributions individually have $\sigma =1$ soft behavior, but the full amplitude has $\sigma = 2$ soft behavior due to nontrivial cancellations between diagrams. For the special galileon, each individual contribution has $\sigma =2$ soft behavior, but their combination has a $\sigma = 3$ soft limit.}
\label{fig:6ptsgal}
\end{figure}
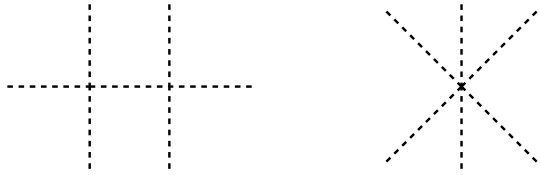

The first nontrivial case is again at six points. There are 10 exchange diagrams that contribute to this amplitude, as depicted in Figure~\ref{fig:6ptsgal}.  The sum of these exchange diagrams has $\sigma=2$ soft behavior, 
\be 
\label{eq:6ptsoft}
\lim_{\tau \rightarrow 0} \mathcal{A}^{\rm exc.}(1_\phi, 2_\phi, 3_\phi, 4_\phi, 5_{\phi}, 6_{\phi}) =\tau^2\left( \frac{2\alpha}{\Lambda^{2(D+2)}} p_{12}\, p_{13}\, p_{23}^2+\cdots\right)+ \cdots,
\ee
where we have only written a representative contribution to the soft limit, as the full expression is rather lengthy. Note that here and in the rest of this section we take the soft limit of the first leg. 
For $D>4$, there is also a six-point contact vertex from the $n=3$ term in Eq.~\eqref{eq:sgal}, as shown in Figure~\ref{fig:6ptsgal}, which has leading soft behavior
\be
\lim_{\tau \rightarrow 0} \mathcal{A}^{\rm cont.}(1_\phi, 2_\phi, 3_\phi, 4_\phi, 5_{\phi}, 6_{\phi}) = \tau^2\left(- \frac{2\alpha}{\Lambda^{2(D+2)}} p_{12}\,p_{13}\,p_{23}^2+\cdots\right)+\cdots.
\ee
This precisely cancels the $\sigma=2$ terms in Eq.~\eqref{eq:6ptsoft}, so the total amplitude has $\sigma =3$ soft behavior. For $D \leq 4$ the six-point term is a total derivative and the exchange diagram by itself has enhanced soft behavior, since the $\tau^2$ terms in Eq.~\eqref{eq:6ptsoft} vanish due to a dimension-dependent Gram identity. 

We can similarly calculate the eight-point special galileon amplitude. This involves exchange diagrams made from the sextic and quartic vertices, exchange diagrams made from three quartic vertices, and a contact term that is non-vanishing in $D >6$. The different diagrams have nontrivial cancellations to achieve the $\sigma =3$ soft behavior of the total amplitude.

\subsection{Minimally coupled free matter}
\label{sec:minimalcoupling}
Having reviewed how amplitudes involving only Goldstone scalars have enhanced soft behavior,
we now verify that this persists in the presence of interactions with matter fields. This should of course be the case, because the interactions have been engineered to preserve the shift symmetries responsible for the soft behavior. Nevertheless, the explicit scattering computation provides a useful consistency check.

\begin{figure}[!t]
\begin{center}
\resizebox{3cm}{!}{
\begin{tikzpicture}[thick, node distance=1.55cm and 1.55cm,line width=1.3 pt]

\coordinate[label= left:$1$] (e1);
\coordinate[below right=of e1] (vertex);
\coordinate[above right=of vertex, label= right:$2$] (e2);
\coordinate[below right=of vertex, label= right:$3$] (e3);
\coordinate[below left=of vertex, label= left:$4$] (e4);

\draw[sgal] (e1) -- (vertex);
\draw[sgal] (e2) -- (vertex);
\draw[scalar] (e3) -- (vertex);
\draw[scalar] (e4) -- (vertex);

\end{tikzpicture}
}
\end{center}
\caption{\small Quartic interaction between the DBI/special galileon (dashed line) and matter (solid line). 
}
\label{fig:quartic}
\end{figure}
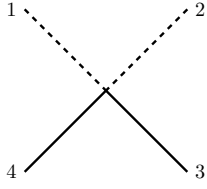

Consider free matter fields interacting with the DBI scalar or special galileon through the minimal couplings introduced in Section~\ref{sec:coupling}. Expanding the determinant and inverse metrics appearing in the kinetic terms leads to an infinite number of vertices, each with two matter fields and an even number of scalar fields.
The lowest-order interaction occurs through a four-point vertex, as depicted in Figure~\ref{fig:quartic}.
For spin-0, spin-1, and massive spin-2 particles, we get the following on-shell four-point amplitudes:
\begin{align}
\mathcal{A}(1_\phi, 2_\phi, 3_\chi, 4_\chi)  =&-\frac{2\alpha}{\Lambda^{(D+2 \kappa)}}  p_{12}^{\kappa} \, p_{13} \,p_{23}, \\
\mathcal{A}(1_\phi, 2_\phi, 3_A, 4_A)  =&-\frac{2\alpha}{\Lambda^{(D+2 \kappa)}} p_{12}^{\kappa} \Big[p_{13}\, p_{23}\, \epsilon_3 \ccdot \epsilon_4+ \left( p_{13} \, \epsilon_3 \ccdot p_2 \, \epsilon_4 \ccdot p_1+ p_{23}  \, \epsilon_3 \ccdot p_1 \, \epsilon_4 \ccdot p_2\right)\Big], \label{eq:4ptspin1}\\
\mathcal{A}(1_\phi, 2_\phi, 3_H, 4_H)  =& -\frac{2\alpha }{\Lambda^{(D+2 \kappa)}} p_{12}^{\kappa} \epsilon_3 \ccdot \epsilon_4 \Big[ \, p_{13}\, p_{23}\, \epsilon_3 \ccdot \epsilon_4+2 \left( p_{13} \, \epsilon_3 \ccdot p_2 \, \epsilon_4 \ccdot p_1+ p_{23} \, \epsilon_3 \ccdot p_1 \, \epsilon_4 \ccdot p_2\right) \nn \\ 
&  +p_{12} \left( \epsilon_3 \ccdot p_2 \, \epsilon_4 \ccdot p_1+\epsilon_3 \ccdot p_1 \, \epsilon_4 \ccdot p_2\right) \Big],
\end{align}
where 
\be \kappa =\begin{cases} 0 & {\rm for\ DBI}, \\ 1 & {\rm for\ the \ special\ galileon}. \end{cases}
\ee
In these amplitudes, the
spin-0 and spin-1 particles can be either massive or massless. It is easily checked that all of these have the desired $\sigma =\kappa+2$ soft behavior. 

In the case of a massless spin-1 particle in four dimensions,  the four-point amplitude \eqref{eq:4ptspin1} with the special galileon precisely matches the amplitude found in Ref.~\cite{Elvang:2018dco} using a soft bootstrap approach. The presence of the same coupling constant in this amplitude and the special galileon four-point amplitude was observed in Ref.~\cite{Elvang:2018dco} and follows from the special galileon version of the equivalence principle, which we discuss in Section~\ref{sec:equiv}.

\begin{figure}[!tb]
\begin{center}
\resizebox{11cm}{!}{
\begin{tikzpicture}[thick, node distance=1.55cm and 1.55cm,line width=1.3 pt]

\coordinate (vertex1);
\coordinate[above= of vertex1] (e1);
\coordinate[left=of vertex1] (e2);
\coordinate[below= of vertex1] (e3);
\coordinate[right=1.5cm of vertex1] (vertex2);
\coordinate[above= of vertex2] (e4);
\coordinate[right =of vertex2] (e5);
\coordinate[below= of vertex2] (e6);
\coordinate[right =.75cm of vertex1] (midpoint1);

\draw[sgal] (e1) -- (vertex1);
\draw[sgal] (e2) -- (vertex1);
\draw[scalar] (e3) -- (vertex1);
\draw[scalar] (vertex1) -- (vertex2);
\draw[sgal] (e4) -- (vertex2);
\draw[sgal] (e5) -- (vertex2);
\draw[scalar] (e6) -- (vertex2);

\coordinate[left=4.5cm of vertex1](vertex3);
\coordinate[above left=2cm of vertex3] (f1);
\coordinate[above=of vertex3] (f2);
\coordinate[above right=2cm of vertex3] (f3);
\coordinate[below right=2cm of vertex3] (f4);
\coordinate[below =of vertex3] (f5);
\coordinate[below left=2cm of vertex3] (f6);

\draw[sgal] (f1) -- (vertex3);
\draw[sgal] (f2) -- (vertex3);
\draw[sgal] (f3) -- (vertex3);
\draw[scalar] (f4) -- (vertex3);
\draw[scalar] (f5) -- (vertex3);
\draw[sgal] (f6) -- (vertex3);

\coordinate[right=4.5cm of vertex2](vertex4);
\coordinate[above= of vertex4] (g1);
\coordinate[left=of vertex4] (g2);
\coordinate[below= of vertex4] (g3);
\coordinate[right=1.5cm of vertex4] (vertex5);
\coordinate[above= of vertex5] (g4);
\coordinate[right =of vertex5] (g5);
\coordinate[below= of vertex5] (g6);

\draw[sgal] (g1) -- (vertex4);
\draw[sgal] (g2) -- (vertex4);
\draw[sgal] (g3) -- (vertex4);
\draw[sgal] (vertex4) -- (vertex5);
\draw[sgal] (g4) -- (vertex5);
\draw[scalar] (g5) -- (vertex5);
\draw[scalar] (g6) -- (vertex5);

\end{tikzpicture}
}
\end{center}
\caption{\small Six-point diagrams with two external matter legs. The number of distinct permutations of each diagram is $(1, 6, 4)$.}
\label{eq:6ptex}
\end{figure}
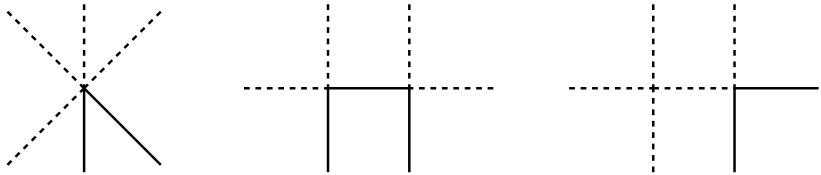

The first non-trivial cancellations again happen at six points. In this case there are 
two amplitudes involving the minimally coupled matter fields. The first has two external matter fields and receives contributions from two types of exchange diagrams, as shown in Figure~\ref{eq:6ptex}.
In each of our examples the sum of the exchange diagrams alone has $\sigma =\kappa+1$ soft behavior, e.g.,
\begin{align}  \label{eq:sixptsoft}
\lim_{\tau \rightarrow 0} \mathcal{A}^{\rm exc.}(1_\phi, 2_\phi, 3_\phi, 4_\phi, 5_{\phi^{(s)}}, 6_{\phi^{(s)}}) = 
\frac{\alpha^2}{\Lambda^{2(D+2\kappa)}} \times \begin{cases}
\tau\left(  -2p_{12} \,p_{23}^2  \left(\epsilon_5 \ccdot \epsilon_6\right)^s + \cdots\right) +\mathcal{O}(\tau^2),  &\kappa=0,\\
\tau^2 \left( 2 p_{14} \,p_{15}\, p_{23}^2 \left(\epsilon_5 \ccdot \epsilon_6\right)^s + \cdots\right) +\mathcal{O}(\tau^3),  &\kappa=1,
\end{cases}
\end{align}
where $\phi^{(s)}$ denotes the spin-$s$ particle, i.e. $\phi^{(0)} = \chi$, $\phi^{(1)} = A$, $\phi^{(2)} = H$.
As before, by expanding out the
inverse metric and determinant in the kinetic terms we also get six-point contact terms, which in the soft limit precisely cancel the $\sigma =\kappa+1$ terms in Eq.~\eqref{eq:sixptsoft}.
The total amplitudes thus have $\sigma =\kappa+2$ soft behavior due to nontrivial cancellations between the Feynman diagrams. 
The other six-point amplitudes are those with four external matter fields, which are built from the four-point matter vertices with the exchange of a scalar. In this case there are no contact terms and each exchange diagram individually has $\sigma=\kappa+2$ soft behavior.

We have also carried out a number of checks of the soft behavior at eight points and verified that all cases have the expected $\sigma = 2+\kappa$ soft behavior.
There are three additional kinds of amplitudes to consider, those with two, four, or six external matter legs. The amplitudes with two external matter legs are built from 336 exchange diagrams (with seven different topologies) plus a contact term, and the amplitudes with four external matter legs are made from 228 exchange diagrams (with four different topologies). 
Each of these amplitudes involve nontrivial cancellations between diagrams to achieve the enhanced $\sigma =\kappa+2$ soft behavior. The remaining class of amplitudes has six external matter legs and is built from 90 exchange diagrams of a single topology. In this case each diagram has $\sigma=2 + \kappa$ soft behavior, so there are no cancellations between diagrams. 

\subsection{Matter self interactions}
\label{sec:chi-soft}
The soft behavior of the amplitudes considered in the previous section is quite nontrivial and requires intricate cancellations between a large number of diagrams. 
A reasonable question is whether there is something special about the case that we considered, where the matter sector is free and all the interactions come from mixing with the DBI or special galileon. 
To investigate this, we consider here two examples where the matter fields have self interactions. 
\subsubsection*{Example 1}
The first example we consider is where the matter field is a a massive scalar with a $\chi^n$ interaction with $n\geq 3$, as given in Eq.~\eqref{eq:scalarselfint}. We consider just the special galileon case for simplicity. Expanding the determinant in this interaction leads to the vertices 
\be
\mathcal{L}_{\chi} \supseteq \frac{1}{n!} \lambda_n  \chi^n- \frac{1}{2n!} \frac{\alpha\lambda_n}{\Lambda^{D+2}} \partial^{\mu} \partial^{\nu} \phi \partial_{\mu} \partial_{\nu} \phi \, \chi^n.
\label{eq:matterints}
\ee
The $n$-point scalar vertex here is $i \lambda_n$ and the $(n+2)$-point vertex is given by
\be \label{eq:n+2contact}
\mathcal{V}\left(1_{\phi}, 2_{\phi}, 3_{\chi}, \dots, (n+2)_{\chi} \right) =  -\frac{i\alpha\lambda_n}{\Lambda^{D+2}}   p_{12}^2.
\ee

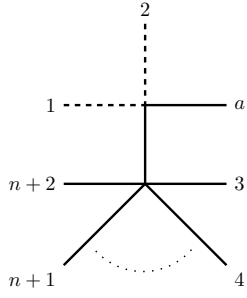
\begin{figure}[!tb]
\begin{center}
\resizebox{3.5cm}{!}{
\begin{tikzpicture}[thick, node distance=1.55cm and 1.55cm,line width=1.3 pt]

\coordinate (vertex1);
\coordinate[left=of vertex1,label=left:$1$] (e1);
\coordinate[above=of vertex1, label=above:$2$] (e2);
\coordinate[right=of vertex1, label=right:$a$] (e3);
\coordinate[below=1.50cm of vertex1] (vertex2);
\coordinate[ right=of vertex2, label=right:$3$] (e4);
\coordinate[left=of vertex2, label=left:$n+2$] (e5);
\coordinate[below left=of vertex2, label=below left:$n+1$] (e6);
\coordinate[below right=of vertex2, label=below right:$4$] (e7);

\draw[loosely dotted, thick] (-.94,-2.78) arc (223:322:1.23cm);
\draw[sgal] (e1) -- (vertex1);
\draw[sgal] (e2) -- (vertex1);
\draw[scalar] (e3) -- (vertex1);
\draw[scalar] (vertex1) -- (vertex2);
\draw[scalar] (e4) -- (vertex2);
\draw[scalar] (e5) -- (vertex2);
\draw[scalar] (e6) -- (vertex2);
\draw[scalar] (e7) -- (vertex2);

\end{tikzpicture}
}
\end{center}
\caption{\small The $(n+2)$-point exchange diagrams in  the $\chi^n$ theory, where $a \in \{3, \dots , n+2 \}$. 
These diagrams alone have $\sigma = 2$ soft behavior, but they combine with the $(n+2)$-point contact term in Eq.~\eqref{eq:matterints} so that the total amplitude has $\sigma = 3$ soft behavior.
}
\label{eq:5ptex}
\end{figure}

Consider the amplitude with two external galileon legs and $n$ matter legs.
By combining the quartic and $n$-point vertices we get the $(n+2)$-point exchange diagrams depicted in Figure~\ref{eq:5ptex}, which contribute to the amplitude as
\be
\mathcal{A}^{\rm exc.}_{n+2}\left(1_{\phi} 2_{\phi} 3_{\chi} \dots (n+2)_{\chi}\right) \Big\rvert_{\lambda_n}=-\frac{\alpha\lambda_n}{\Lambda^{D+2}} \, p_{12} \sum_{a=3}^{n+2} \frac{ p_{1a} \,p_{2a}}{p_{12} +p_{1a}+p_{2a}} \,,
\ee
where the notation $~\rvert_{\lambda_n}$ means that we only write the part of the amplitude proportional to $\lambda_n$.
In the soft limit this gives
\be \label{eq:5ptsoft}
\lim_{\tau \rightarrow 0}\mathcal{A}^{\rm exc.}_{n+2}\left(1_{\phi} 2_{\phi} 3_{\chi} \dots (n+2)_{\chi}\right)\Big\rvert_{\lambda_n}  =  \tau^2\frac{\alpha\lambda_n}{\Lambda^{D+2}} p_{12}^2 + \mathcal{O}(\tau^3),
\ee
which has $\sigma=2$ soft behavior. This is precisely cancelled by the contribution from the contact term \eqref{eq:n+2contact}, confirming that the part of this amplitude proportional to $\lambda_n$ has $\sigma=3$ soft behavior, as required by the symmetry. 

\subsubsection*{Example 2}
As another example, consider the coupling of the special galileon to a cubic galileon, i.e., the interaction $ \lambda (\partial \chi)^2 \Box \chi$ with coupling $\lambda$, where $m_{\chi} =0$ and indices are contracted using $\bar{g}_{\mu \nu}$, as in Eq.~\eqref{eq:scalarselfint}. Expanding the metric gives the cubic and quintic interactions,
\begin{align}
\mathcal{L}_{\chi} \supseteq &\frac{\alpha\lambda}{\Lambda^{D+2}} \left[ (\partial \chi)^2\left( \frac{1}{2}\partial^{\alpha} \partial^{\beta} \phi \partial_{\alpha} \partial_{\beta} \phi \partial^{\mu} -\partial^{\mu} \partial^{\alpha} \phi \partial_{\alpha} \partial^{\nu} \phi\partial_{\nu} -\partial^{\mu} \partial_{\nu} \phi \partial^{\nu} \Box \phi \right) - \partial^{\nu} \chi \Box \chi \partial^{\mu} \partial^{\alpha} \phi \partial_{\alpha} \partial_{\nu} \phi  \right] \partial_{\mu}\chi \nn \\
&+ \lambda (\partial \chi)^2 \Box \chi,
\end{align}
where the quintic interactions receive contributions from the connection of the effective metric. The cubic vertex and the quintic vertex with all but the fifth leg on-shell are given by
\begin{align}
\mathcal{V}\left(1_{\chi}, 2_{\chi}, 3_{\chi} \right) &= 2 i \lambda \left(p_{12}\, p_{33}+p_{13}\, p_{22} + p_{23}\, p_{11} \right), \label{eq:gal-vertex} \\
\mathcal{V}\left(1_{\phi}, 2_{\phi}, 3_{\chi}, 4_{\chi}, 5_{\chi} \right)& = -\frac{2 i \alpha \lambda}{\Lambda^{D+2}} p_{12} \Big[ 2\left(p_{13}\, p_{23} \,p_{45} + p_{14} \,p_{24} \,p_{35}+p_{15}\, p_{25}\, p_{34}\right) \nn \\
&\quad \, +p_{55}\left(p_{14} p_{23} +p_{13} p_{24}-p_{12} p_{34} \right) \Big]. \label{eq:quintic-vertex}
\end{align}

\begin{figure}[!tb]
\begin{center}
\resizebox{11cm}{!}{
\begin{tikzpicture}[thick, node distance=1.55cm and 1.55cm,line width=1.3 pt]

\coordinate (vertex1);
\coordinate[above left= of vertex1] (e1);
\coordinate[below left=of vertex1] (e2);
\coordinate[right= 1cm of vertex1] (e3);
\coordinate[above=of e3] (e4);
\coordinate[below= of e3] (e5);
\coordinate[right = 1cm of e3] (vertex2);
\coordinate[below right= of vertex2] (e6);
\coordinate[above right = of vertex2] (e7);

\draw[scalar] (e1) -- (vertex1);
\draw[scalar] (e2) -- (vertex1);
\draw[scalar] (e3) -- (vertex1);
\draw[sgal] (e3) -- (e5);
\draw[sgal] (e3) -- (e4);
\draw[scalar] (e3) -- (vertex2);
\draw[scalar] (e6) -- (vertex2);
\draw[scalar] (e7) -- (vertex2);

\coordinate[left=8cm of e3] (vertex3);
\coordinate[above left= of vertex3] (f1);
\coordinate[below left=of vertex3] (f2);
\coordinate[right= 1cm of vertex3] (f3);
\coordinate[above=of f3] (f4);
\coordinate[left= 1.7cm of vertex3] (f5);
\coordinate[right = 1cm of f3] (vertex4);
\coordinate[below right= of vertex4] (f6);
\coordinate[above right = of vertex4] (f7);

\draw[sgal] (f1) -- (vertex3);
\draw[sgal] (f2) -- (vertex3);
\draw[scalar] (f3) -- (vertex3);
\draw[scalar] (f3) -- (f5);
\draw[scalar] (f3) -- (f4);
\draw[scalar] (f3) -- (vertex4);
\draw[scalar] (f6) -- (vertex4);
\draw[scalar] (f7) -- (vertex4);

\coordinate[right=6cm of e3] (vertex5);
\coordinate[above left= of vertex5] (g1);
\coordinate[below left=of vertex5] (g2);
\coordinate[above= of vertex5] (g3);
\coordinate[below= of vertex5] (g4);
\coordinate[right= 1.3cm of vertex5] (vertex6);
\coordinate[below right= of vertex6] (g5);
\coordinate[above right = of vertex6] (g6);

\draw[scalar] (g1) -- (vertex5);
\draw[scalar] (g2) -- (vertex5);
\draw[sgal] (g3) -- (vertex5);
\draw[sgal] (g4) -- (vertex5);
\draw[scalar] (vertex5) -- (vertex6);
\draw[scalar] (g5) -- (vertex6);
\draw[scalar] (g6) -- (vertex6);

\end{tikzpicture}
}
\end{center}
\caption{\small Six-point diagrams with two external special galileons and four galileons. The number of distinct permutations of each diagram is $( 12, 3, 6)$.}
\label{eq:6ptex2}
\end{figure}
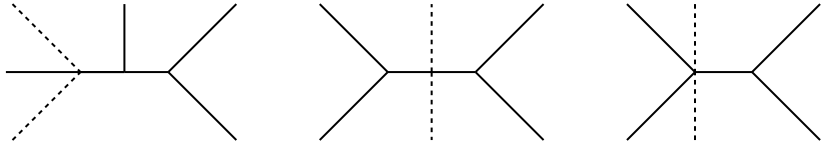

Consider the six-point amplitude with two external special galileon legs and four regular galileon legs (the five-point amplitude vanishes identically). This amplitude recieves contributions from 15 exchange diagrams built from the minimal coupling quartic vertex and two cubic vertices (with two different topologies) and from six exchange diagrams built from the cubic and quintic vertices, as shown in Fig.~\ref{eq:6ptex2}. The leading single soft behavior of the first set of these exchange diagrams is
\be
\lim_{\tau \rightarrow 0}\mathcal{A}^{\rm exc. \, 1}_{6}\left(1_{\phi} 2_{\phi} 3_{\chi} 4_{\chi} 5_{\chi} 6_{\chi} \right)\Big\rvert_{\lambda}  =  \tau^2 \frac{\alpha \lambda^2 }{\Lambda^{D+2}} \left( -24 p_{12}\, p_{16}\, p_{23}^2\, p_{24} + \dots \right) + \mathcal{O}(\tau^3),
\ee
and the leading single soft behavior of the second set of exchange diagrams is
\be 
\lim_{\tau \rightarrow 0}\mathcal{A}^{\rm exc. \, 2}_{6}\left(1_{\phi} 2_{\phi} 3_{\chi} 4_{\chi} 5_{\chi} 6_{\chi} \right)\Big\rvert_{\lambda}  =  \tau^2 \frac{\alpha \lambda^2 }{\Lambda^{D+2}} \left( 24 p_{12}\, p_{16} \,p_{23}^2\, p_{24} + \dots \right) + \mathcal{O}(\tau^3).
\ee
These precisely cancel at order $\tau^2$, so the total amplitude has $\sigma =3$ single soft behavior on the first leg. This demonstrates how higher-derivative matter interactions can give the expected soft behavior through nontrivial cancellations between diagrams. 

\section{Soft scalar equivalence principles}
\label{sec:equiv}

Gravity is universal. This universality can, in a sense, be thought of as a defining property of GR. Indeed, it was taking the equivalence principle seriously that led Einstein to construct GR. The apparatus of $S$-matrix theory provides an interesting twist to the logic---from this viewpoint the equivalence principle is an {\it output}, following from simultaneously demanding locality and Lorentz invariance for massless spin-2 particles.
In this Section, we will show how the soft scalar EFTs have a precise analogue of the equivalence principle, deepening their analogy with gravity.

It is worth first briefly reviewing how the gravitational equivalence principle manifests in the $S$-matrix approach. 
A bosonic spin-$s$ particle, $\phi^{(s)}_{\mu_1 \dots \mu_s}$, which is coupled to the graviton, $h_{\mu \nu}$, or which couples to any particle that couples to gravity, must interact with gravity through the following on-shell cubic vertex
\be \label{eq:gr-min}
\mathcal{V}\big(1_h, 2_{\phi^{(s)}}, 3_{\phi^{(s)}} \big) = -\frac{2i}{M_{\rm Pl}^{\frac{D-2}{2}}}( \epsilon_1 \cdot p_2) \, ( \epsilon_1 \cdot p_3) \, (\epsilon_2 \cdot \epsilon_3)^s + \mathcal{O}(p_1) \, ,
\ee
where the coupling is universal and set by the reduced Planck mass, $M_{\rm Pl}$, and where any other contributions to this vertex are at least linear in the graviton momentum $p_1$.
This is the $S$-matrix equivalence principle.  Weinberg proved this using purely on-shell arguments by imposing the on-shell Ward identity on scattering amplitudes with a soft graviton leg~\cite{Weinberg:1964ew}. This result relies on the technical assumption that the set of graviton cubic interactions contains the cubic vertex of GR, in addition to the usual assumptions of Lorentz invariance and locality. 
In the Lagrangian approach, this part of the vertex arises from minimally coupling a particle's kinetic term to the metric, so we refer to it as the minimal coupling vertex.

Both DBI scalars and the special galileon obey an interesting analogue of the equivalence principle. One might suspect that something like this should be true because there is a covariant effective metric that all matter fields can couple to in a way that preserves the shift symmetry.
In this section, we put this Lagrangian intuition on-shell by deriving the DBI and special galileon versions of the $S$-matrix equivalence principle. We assume as part of their definition that these theories have a $\mathbb{Z}_2$ symmetry under $\phi \mapsto - \phi$.

\subsection{DBI equivalence principle}
\label{sec:DBIequiv}

The statement we will derive is quite similar to the usual gravitational equivalence principle. Any particle that is coupled to a DBI scalar---or any particle that couples to a particle that couples to a DBI scalar---must interact with the DBI scalar through an on-shell quartic vertex with the following form:
\be \label{eq:dbi-min}
\mathcal{V} \big( 1_\phi, 2_\phi, 3_{\phi^{(s)}},4_{\phi^{(s)}}\big) = -\frac{2i \alpha }{\Lambda^D} p_{13} \,p_{23}\, (\epsilon_3\cdot\epsilon_4)^s+\mathcal{O}(\xi^3),
\ee
where the universal coupling $\alpha$ is the DBI coupling and $\xi$ counts the total combined power of $p_1$ and $p_2$.
This leading interaction arises in the Lagrangian approach from minimally coupling a particle's kinetic term to the DBI metric, so we refer to it as the minimal coupling vertex.

\subsubsection*{On-shell proof}
\label{sec:dbiequiv}
We now show from the on-shell perspective why such couplings must be universal. The derivation parallels the derivation of the gravitational $S$-matrix equivalence principle~\cite{Weinberg:1964ew}.
In Weinberg's derivation, there are two crucial components: the universality of the leading interactions in the single soft limit and the on-shell Ward identity.
We will see that the analogue of the single soft limit of the graviton is the double soft limit of DBI scalars and the analogue of the on-shell Ward identity is the vanishing single soft theorem.

To begin, consider a general $N$-point amplitude, $\mathcal{A}_N$, with $N>2$ and where the $a$\textsuperscript{th} particle has spin $s_a$, mass $m_a$, and a polarization tensor $\epsilon^{(a)}_{\mu_1 \dots \mu_{s_a}}$. 
It will sometimes be helpful to strip off the polarization of the $a$\textsuperscript{th} particle and define $\mathcal{A}_N = \epsilon^{(a)}_{\mu_1 \dots \mu_{s_a}} \mathcal{A}_{N, a}^{\mu_1 \dots \mu_{s_a}}(p_a)$, where the dependence of the amplitude on the other momenta is suppressed. 
We then attach to $\mathcal{A}_N$ two DBI legs with momenta $p_{N+1}$ and $p_{N+2}$ and denote the resulting amplitude by $\mathcal{A}_{N+2}$. After scaling $p_{N+1} \rightarrow \xi p_{N+1}$ and $p_{N+2} \rightarrow \xi p_{N+2}$, we want to find the leading part of $\mathcal{A}_{N+2}$ in the limit $\xi\rightarrow 0$, i.e., the double soft limit.\footnote{One can consider other double soft limits, where the different legs are taken soft at different rates, but we do not need these.} Note that to make this procedure unambiguous we use momentum conservation to maximize the smallest exponent of $\xi$ appearing in the double soft limit of $\mathcal{A}_{N+2}$, i.e. the double soft degree of $\mathcal{A}_{N+2}$. This can be achieved, for example, by eliminating all occurrences of any one momentum that is not taken soft plus one additional contraction not involving the soft momenta. 

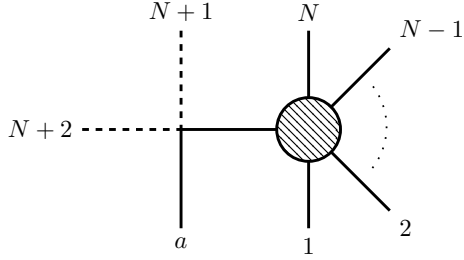
\begin{figure}[!t]
\begin{center}
\resizebox{6.5cm}{!}{
\begin{tikzpicture}[thick, node distance=1.55cm and 1.55cm,line width=1.3 pt]

\coordinate (vertex1);
\coordinate[above=of vertex1, label=above:$N+1$] (e1);
\coordinate[left=of vertex1, label=left:$N+2$] (e2);
\coordinate[below=of vertex1, label=below:$a$] (e3);
\coordinate[right=2cm of vertex1] (vertex2);
\coordinate[below=of vertex2, label=below:$1$] (e4);
\coordinate[below right=1.8cm of vertex2, label=below right:$2$] (e5);
\coordinate[above right=1.8cm of vertex2, label=above right:$N-1$] (e6);
\coordinate[above=of vertex2, label=above:$N$] (e7);

\draw[loosely dotted, thick] (3,-.605) arc [start angle=-38, end angle=38, radius=1.05cm];
\draw[sgal] (e1) -- (vertex1);
\draw[sgal] (e2) -- (vertex1);
\draw[scalar] (e3) -- (vertex1);
\draw[scalar] (vertex1) -- (vertex2);
\draw[scalar] (e4) -- (vertex2);
\draw[scalar] (e5) -- (vertex2);
\draw[scalar] (e6) -- (vertex2);
\draw[scalar] (e7) -- (vertex2);
\fill[white] (2,0) circle (.5cm);
\draw[pattern=north west lines] (2,0) circle (.5cm);

\end{tikzpicture}
}
\end{center}
\caption{\small The leading diagrams in the double soft limit of $\mathcal{A}_{N+2}$. The dashed lines are the soft scalar legs and the solid lines denote arbitrary external particles.}
\label{fig:doublesoft}
\end{figure}

The leading contributions to the double soft limit come from exchange diagrams where the two scalar legs meet at a quartic vertex on an external line of $\mathcal{A}_N$,  as shown in Figure~\ref{fig:doublesoft}. Only these diagrams can contribute pole terms in the double soft limit. Any other exchange diagrams will give subleading contributions in the double soft limit and can thus be ignored. Contact diagrams can also be ignored since they do not contribute at leading order in the double soft limit, as we explain more fully below.

Moreover, we only need to consider quartic vertices that have the required DBI single soft behavior and the minimal double soft degree for the first two legs. It turns out that there is a unique such vertex, as we explain in Appendix~\ref{app:4ptverts}, which has double soft degree two and takes the form of the minimal coupling vertex~\eqref{eq:dbi-min},
\be \label{eq:DBI-leading-quartic}
\mathcal{V} \big( 1_\phi, 2_\phi, 3_{\phi^{(s_a)}}, 4_{\phi^{(s_a)}}\big) = -\frac{2i \alpha_a }{\Lambda^D} p_{13} \,p_{23}\, (\epsilon_3\cdot\epsilon_4)^{s_a},
\ee
where $\alpha_a$ is the coupling constant for the $a$\textsuperscript{th} particle in the amplitude $\mathcal{A}_N$, which a priori can take any value. If particle $a$ is itself a DBI scalar, then from Eq.~\eqref{eq:DBIquartic} we have that $\alpha_a = \alpha$.

We can now calculate the leading contribution to the double soft limit of $\mathcal{A}_{N+2}$ by summing over the diagrams of the type shown in Figure~\ref{fig:doublesoft} using the quartic vertices \eqref{eq:DBI-leading-quartic},
\be \label{eq:DBIdouble-soft-1}
\lim_{\xi \rightarrow 0} \mathcal{A}_{N+2}  = -\lim_{\xi \rightarrow 0}2 \xi^2  \sum_{a=1}^{N} \frac{\alpha_a}{\Lambda^{D}} p_{N+1} \ccdot p_{a}\, p_{N+2} \ccdot p_{a} \, \epsilon^{(a)}_{\lambda_1 \dots \lambda_{s_a}} \frac{\Pi^{\lambda_1 \dots \lambda_{s_a}}{}{}_{ \mu_1 \dots \mu_{s_a}}(\tilde{p}_a)}{\tilde{p}_a^2+m_a^2}\mathcal{A}^{\mu_1 \dots \mu_{s_a}}_{N, a}(\tilde{p}_a)+ \mathcal{O}(\xi^2),
\ee
where we have defined the shifted momentum $\tilde{p}_a \equiv p_a+\xi p_{N+1}+\xi p_{N+2}$ and $\Pi_{\nu_1 \dots \nu_{s_a}, \mu_1 \dots \mu_{s_a}}$ is the numerator of the propagator for particle $a$. The factors of $\xi$ in the propagator denominator mean that there are ${\cal O}(\xi)$ contributions to Eq.~\eqref{eq:DBIdouble-soft-1}.  We can simplify this expression using the completeness relation for the on-shell propagator,
\be
\Pi_{\nu_1 \dots \nu_{s_a}, \mu_1 \dots \mu_{s_a}} = \sum_{\ell} \epsilon^*{}^{(\ell)}_{\nu_1 \dots \nu_{s_a}}\epsilon^{(\ell)}_{\mu_1 \dots \mu_{s_a}} + \dots ,
\ee
where $\ell$ runs over the polarization states of the intermediate particle and the ellipsis denotes gauge-dependent longitudinal terms that are present for massless particles but that drop out in Eq.~\eqref{eq:DBIdouble-soft-1} because they contract with the polarization and the gauge invariant amplitude. 
This gives
\be \label{eq:dbipoleformula}
\lim_{\xi \rightarrow 0} \mathcal{A}_{N+2}  = -\xi \sum_{a=1}^{N}\frac{\alpha_a}{\Lambda^{D}} \frac{ p_{N+1} \ccdot p_{a}\, p_{N+2} \ccdot p_{a}}{ (p_{N+1}+p_{N+2}) \ccdot p_{a}}\mathcal{A}_{N}+ \mathcal{O}(\xi^2).
\ee

Equation~\eqref{eq:dbipoleformula} follows from general principles---essentially just $S$-matrix factorization along with the input that the interaction takes the form~\eqref{eq:DBI-leading-quartic}. To make further progress, we now add nontrivial input from the DBI theory, namely that it has enhanced single soft behavior. We
demand that $\mathcal{A}_{N+2}$ has $\sigma=2$ single soft behavior on the DBI legs, which we impose by scaling $p_{N+1}$ to zero faster than $p_{N+2}$ in the formula~\eqref{eq:dbipoleformula}. Importantly, this condition must
hold order-by-order in the double soft parameter $\xi$. Rescaling $p_{N+1} \rightarrow \tau p_{N+1}$ and then taking $\tau$ to zero gives
\be \label{eq:softsoft}
\lim_{\tau \rightarrow 0} \lim_{\xi \rightarrow 0} \mathcal{A}_{N+2} =  -\frac{\tau\, \xi }{\Lambda^{D}}\mathcal{A}_{N} \, p_{N+1} \cdot \left(\sum_{a=1}^{N}\alpha_a p_{a}\right)+ \mathcal{O}( \tau^2,\xi)+ \mathcal{O}(\xi^2).
\ee
The first term is $\mathcal{O}(\tau)$ and must therefore vanish or cancel against other terms for the amplitude to have $\sigma=2$ single soft behavior.
A local contact term cannot contribute at $\mathcal{O}(\xi)$ since these must be invariant under $p_{N+1} \leftrightarrow p_{N+2}$ and so only contribute even powers of $\xi$ in the double soft limit when $N>2$. {There can also be no $\mathcal{O}(\xi^0)$ contact terms since these would be inconsistent with the required single soft behavior.} 
It is then clear that the only way to have the desired single soft behavior is if the couplings $\alpha_a$ are all equal, since then the leading term is proportional to $p_{N+1} \ccdot p_{N+2}\,\mathcal{A}_{N}$ by momentum conservation, which is $\mathcal{O}(\xi^2)$ and can cancel against other terms.\footnote{This is a necessary but not sufficient condition for the total amplitude to have the requisite soft behavior, which is completely parallel to the gravitational $S$-matrix equivalence principle, which is a necessary but not sufficient conditions for the full amplitude to be gauge invariant.}
Moreover, the universal coupling must be equal to $\alpha$, the DBI self-coupling, since we can consider the case when $\mathcal{A}_N$ has an external DBI scalar.

We thus conclude that all particles must either couple to the DBI scalar through the minimal coupling quartic vertex with the same universal coupling constant or else completely decouple.

\subsection{Special galileon equivalence principle}
\label{sec:sgalequiv}
The DBI $S$-matrix equivalence principle derived above also has a special galileon analogue. In this case the statement is that particles must couple to the special galileon through
the following on-shell quartic vertex:
\be \label{eq:sgal-min}
\mathcal{V} \big( 1_\phi, 2_\phi, 3_{\phi^{(s)}}, 4_{\phi^{(s)}}\big) =- \frac{2i \alpha}{\Lambda^{D+2}} p_{12}\, p_{13}\, p_{23} \,(\epsilon_3 \cdot \epsilon_4)^s + \mathcal{O}(\xi^{5})\, ,
\ee
where the universal coupling is given by $\alpha$ and where $\xi$ again counts the total combined power of $p_1$ and $p_2$, or else they must completely decouple from anything that couples to the special galileon.
This vertex  follows from minimal coupling of the matter kinetic term with the covariant metric~\eqref{eq:sgalmetric}, so we refer to this as the minimal coupling vertex. In addition to non-minimal interactions involving curvature tensors of the metric, there are invariant interactions that can be built from the other covariant building blocks of the coset construction~\cite{Garcia-Saenz:2019yok,Carrillo-Gonzalez:2019aao}, but these do not modify the vertex \eqref{eq:sgal-min}.

\subsubsection*{On-shell proof}

We can prove the special galileon equivalence principle in a similar way to the DBI case. Since many of the algebraic manipulations are similar, we will be more telegraphic in this derivation. The essential ingredients are again the special galileon double soft limit and the vanishing single soft theorem.

Consider adding two additional special galileon legs with momenta $p_{N+1}$ and $p_{N+2}$ to an $N$-point amplitude $\mathcal{A}_N = \epsilon^{(a)}_{\mu_1 \dots \mu_{s_a}} \mathcal{A}_{N, a}^{\mu_1 \dots \mu_{s_a}}(p_a)$, rescaling
$p_{N+1} \rightarrow \xi p_{N+1}$ and $p_{N+2} \rightarrow \xi p_{N+2}$, and taking the double soft limit $\xi\to 0$. The leading diagrams in this limit are again the exchange diagrams depicted in Figure~\ref{fig:doublesoft}. The minimal double soft degree of a quartic vertex with the required $\sigma=3$ single soft behavior is four. An example of such a vertex is that with the structure of the minimal coupling vertex, 
\be \label{eq:sgal-min-gen}
\mathcal{V} \big( 1_\phi, 2_\phi, 3_{\phi^{(s_a)}}, 4_{\phi^{(s_a)}}\big) =- \frac{2i \alpha_a}{\Lambda^{D+2}} p_{12}\, p_{13}\, p_{23} \,(\epsilon_3 \cdot \epsilon_4)^{s_a} \, ,
\ee
where $\alpha_a$ is the coupling constant for the $a$\textsuperscript{th} particle in the amplitude $\mathcal{A}_N$. This coupling is $\alpha$ if particle $a$ is itself a special galileon.
Unlike in the DBI argument, there are multiple vertices with the same double soft degree as the minimal coupling vertex, so these all contribute at the same order in the double soft limit. It turns out that these other vertices must be absent to be consistent with the $\sigma=3$ single soft behavior, as we explain in detail in Appendix~\ref{app:4ptverts}, so we again only have to consider the vertex \eqref{eq:sgal-min-gen}. 

We can now calculate the leading contribution to the double soft limit of $\mathcal{A}_{N+2}$ by summing over the diagrams in Figure~\ref{fig:doublesoft} using the quartic vertex \eqref{eq:sgal-min-gen}.
Following the same steps as for DBI gives
\be \label{eq:poleformula}
\lim_{\xi \rightarrow 0} \mathcal{A}_{N+2}  =- \xi^3\,p_{N+1} \ccdot p_{N+2} \sum_{a=1}^{N}\frac{\alpha_a}{\Lambda^{D+2}} \frac{p_{N+1} \ccdot p_{a}\,p_{N+2} \ccdot p_{a}}{(p_{N+1}+p_{N+2}) \ccdot p_{a}}\mathcal{A}_{N}+ \mathcal{O}(\xi^4).
\ee
We now demand that $\mathcal{A}_{N+2}$ has $\sigma=3$ single soft behavior.  Rescaling $p_{N+1}\rightarrow \tau p_{N+1}$ and then taking $\tau$ to zero gives
\be \label{eq:softsoft}
\lim_{\tau \rightarrow 0} \lim_{\xi \rightarrow 0} \mathcal{A}_{N+2} = - \frac{\tau^2 \xi^3}{\Lambda^{D+2}}p_{N+1} \ccdot p_{N+2} \mathcal{A}_{N} \, p_{N+1} \cdot \left(\sum_{a=1}^{N}\alpha_a p_{a}\right)+ \mathcal{O}( \tau^3,\xi^3)+ \mathcal{O}(\xi^4).
\ee
The first term is $\mathcal{O}(\tau^2)$ and must therefore vanish or cancel against other terms for $\mathcal{A}_{N+2}$ to have $\sigma=3$ single soft behavior. Contact terms cannot cancel this term since they have an even double soft degree for $N>2$. The only way to achieve this is if the coupling constants are all equal, since then the leading term is proportional to $(p_{N+1} \ccdot p_{N+2})^2 \mathcal{A}_N $ by momentum conservation, which is of order $\xi^4$ and can thus cancel against other terms. Moreover, we must have $\alpha_a =\alpha$, the special galileon self coupling, since $\mathcal{A}_N$ can have an external special galileon leg.

This implies that 
all particles must couple to the special galileon through the minimal coupling quartic vertex with coupling constant $\alpha$ or else completely decouple. Note that this argument does not go through for a generic galileon theory, since these only have $\sigma=2$ single soft behavior, which does not constrain the couplings $\alpha_a$.

\subsection{No massless higher-spin interactions}

Now that we have derived the DBI and special galileon equivalence principles, we can utilize them to prove some further interesting facts. For example, it is not possible to consistently couple these scalars to gravity or massless higher-spin particles, since their equivalence principles are inconsistent with higher-spin gauge invariance.
This echoes the fact that
when $s>2$, the gravitational minimal coupling vertex \eqref{eq:gr-min} is incompatible with the higher-spin gauge invariance that would be required for a massless spin-$s$ particle, implying that gravity cannot couple to massless higher-spin fields in flat space~\cite{Aragone:1979hx, Metsaev:2005ar, Porrati:2008rm, Boulanger:2008tg, Taronna:2011kt}.

Consider the minimal coupling quartic vertices \eqref{eq:dbi-min} and \eqref{eq:sgal-min} between the soft scalars and a massless spin-$s$ particle. Under a gauge variation of the spin-$s$ field, where we shift $\epsilon_3\mapsto \epsilon_3 + \varepsilon p_3$, these vertices change at linear order in $\varepsilon$ by
\be \label{eq:deltamin}
\delta \mathcal{V}\big( 1_\phi, 2_\phi, 3_{\phi^{(s)}},4_{\phi^{(s)}}\big) = -\frac{2 i \alpha}{\Lambda^{D+1}} s \, p_{12}^{\kappa} \,p_{13}\, p_{23} (\epsilon_4 \cdot p_3 )(\epsilon_3 \cdot \epsilon_4)^{s-1}+\mathcal{O}(\xi^{2\kappa+3}),
\ee
where $\kappa =0$ for the DBI scalar, $\kappa = 1$ for the special galileon, and $\xi$ again counts the total power of $p_1$ and $p_2$.
This must be cancelled by the gauge variation of other terms if the spin-$s$ particle is massless.
Any additional terms must have $2\kappa+4$ derivatives if their gauge variations are to help cancel \eqref{eq:deltamin}, since the gauge variation of a vertex preserves the number of derivatives when all particles are massless. A direct check of all possibilities shows that no vertices with the requisite properties exist except when $s=1$, as we now describe. This is the soft scalar analogue of the generalized Weinberg--Witten theorem \cite{Metsaev:2005ar, Porrati:2008rm,Boulanger:2008tg}.

We begin by enumerating all possible
on-shell quartic vertices with $2\kappa+4$ derivatives and which vanish in the soft limit with $\sigma \geq \kappa+2$. The result is a sum of terms of the form 
\be \label{eq:genvertex1}
 p_{12}^{r_{12}}\, p_{13}^{r_{13}} (\epsilon_3 \ccdot \epsilon_4)^{n_{34}} (\epsilon_3 \ccdot p_1)^{m_{31}}   (\epsilon_3 \ccdot p_2)^{m_{32}} (\epsilon_4 \ccdot p_1)^{m_{41}} (\epsilon_4 \ccdot p_2)^{m_{42}},
\ee
where the exponents are non-negative integers satisfying the linear system of equations
\begin{subequations}
\label{eq:linearcons1}
\begin{align}
n_{34} + m_{31} +m_{32} & = s, \\
n_{34}+ m_{41} + m_{42} & =s, \\
m_{31}+m_{41}+r_{12}+r_{13} & \geq \kappa+2, \label{eq:ineq11} \\
m_{32}+m_{42}+r_{12}+r_{13} & \geq \kappa+2, \label{eq:ineq21}\\
m_{31} +m_{32}+ m_{41} + m_{42}+2r_{12}+2r_{13} &=2\kappa + 4.
\end{align}
\end{subequations}
These conditions ensure that the two particles have the same spin and that the vertex has $2\kappa+4$ derivatives and  is consistent with having $\sigma = \kappa+2$ soft behavior.

It is straightforward to explicitly find all solutions to these equations for each spin and to further impose that they are the on-shell part of a vertex with the correct particle interchange symmetries. This leaves a $(4\kappa+6)$-parameter family of vertices for each $s>2$ (there can be fewer for $s \leq 2$), including the minimal coupling vertex. 

Imposing the on-shell Ward identity for the higher-spin leg, we find that there are no gauge invariant vertices amongst these general families for $s \geq 2$. For $s=1$, there are unique gauge-invariant completions of the minimal coupling vertices,
\be
{\cal V}\big( 1_\phi, 2_\phi, 3_A, 4_A \big)=-\frac{2i\alpha}{\Lambda^{D+2 \kappa}}p_{12}^{\kappa}\Big[p_{13}\, p_{23}\, \epsilon_3 \ccdot \epsilon_4+ \left( p_{13} \, \epsilon_3 \ccdot p_2 \, \epsilon_4 \ccdot p_1+ p_{23}  \, \epsilon_3 \ccdot p_1 \, \epsilon_4 \ccdot p_2\right)\Big] \, ,
\ee
which are the full on-shell vertices for a photon minimally coupled to a DBI scalar or special galileon. 
The absence of solutions for $s\geq 2$, combined with the equivalence principles, shows that the DBI scalar and the special galileon, like gravity, cannot couple to massless higher-spin fields, including gravity. A possible loophole is if the higher-spin interactions take a noncovariant form, as for the light-cone vertices used in some proposed four-dimensional flat space higher-spin theories \cite{Metsaev:1991mt, Metsaev:1991nb, Ponomarev:2016lrm, Skvortsov:2018jea} (see also Refs.~\cite{Sleight:2016xqq, Taronna:2017wbx}). Another possible loophole is for parity-odd theories in $D \leq 5$, which we have not considered here.

\subsection{Double soft theorems}
\label{sec:doublesoft}
A byproduct of the proofs of the soft scalar equivalence principles is that the leading double soft limits of DBI and special galileon matter amplitudes have a universal factorized form,
\be \label{eq:soft-scalar-factorization}
\lim_{\xi \rightarrow 0} \mathcal{A}_{N+2}  = -\xi^{2\kappa+1} \frac{\alpha}{\Lambda^{D+2\kappa}} (p_{N+1} \ccdot p_{N+2})^{\kappa} \mathcal{A}_{N}\sum_{a=1}^{N} \frac{ p_{N+1} \ccdot p_{a}\, p_{N+2} \ccdot p_{a}}{ (p_{N+1}+p_{N+2}) \ccdot p_{a}}+ \mathcal{O}(\xi^{2\kappa+2}).
\ee
These are the analogues of Weinberg's soft graviton theorem~\cite{Weinberg:1964ew, Weinberg:1965nx} and are well-known for the pure scalar theories~\cite{Cachazo:2015ksa,Li:2017fsb,Guerrieri:2017ujb},\footnote{The analogy with the graviton soft theorem was already pointed out in Ref.~\cite{Cachazo:2015ksa}.} but here we see that they hold also in the presence of matter. 

For the graviton, the leading soft theorem is just the beginning of the story, and the graviton satisfies also a subleading soft theorem~\cite{Cachazo:2014fwa,Schwab:2014xua,Campiglia:2014yka}.\footnote{There are also interesting connections between soft theorems, asymptotic symmetries, and memory effects, as reviewed in Ref.~\cite{Strominger:2017zoo}. Scalar analogues of these relations have been studied in, e.g., Refs.~\cite{Campiglia:2017dpg, Hamada:2017atr, Campiglia:2017xkp}. It would be interesting to further explore these connections for the theories considered here.} Similarly, there are universal subleading terms in the double soft expansions of the pure scalar theories~\cite{Cachazo:2015ksa,Li:2017fsb,Guerrieri:2017ujb}, so it is interesting to explore whether or not these continue to hold in the presence of matter. Here we briefly review these subleading results and discuss how they generalize to amplitudes involving matter fields. 

\subsubsection*{DBI and special galileon double soft theorems}

We begin by reviewing the statements of the double soft theorems for the DBI scalar and special galileon, which were originally discovered in Ref.~\cite{Cachazo:2015ksa}.
If we scale
$p_{N+1} \rightarrow \xi p_{N+1}$ and $p_{N+2} \rightarrow \xi p_{N+2}$ in an $(N+2)$-point amplitude $\mathcal{A}_{N+2}$ with all DBI legs ($\kappa=0$) or all special galileon legs ($\kappa=1$)  and take $\xi \rightarrow 0$, the amplitude factorizes into a series of soft factors times the $N$-point amplitude obtained by removing the two soft legs~\cite{Cachazo:2015ksa}:
\be \label{eq:doublesoft}
\lim_{\xi \rightarrow 0} \mathcal{A}_{N+2}  =\frac{\alpha}{\Lambda^{D+2 \kappa}}\,\xi^{2\kappa} (p_{N+1} \ccdot p_{N+2})^{\kappa} \sum_{j=0}^2 S^{(j)} \mathcal{A}_{N}+ \mathcal{O}(\xi^{2\kappa+4}),
\ee
where the various soft factors are given by
\begin{align}
S^{(0)} & = \frac{\xi}{4} \sum_{a=1}^{N} \left( \frac{(p_{N+1} \ccdot p_a-p_{N+2} \ccdot p_a)^2}{(p_{N+1}+p_{N+2} ) \ccdot p_a+\xi p_{N+1} \ccdot p_{N+2}}+(p_{N+1}+p_{N+2} ) \ccdot p_a+\xi p_{N+1} \ccdot p_{N+2}\right), \\
S^{(1)} & = \frac{\xi^2}{2} \sum_{a=1}^{N}  \frac{(p_{N+1}-p_{N+2} ) \ccdot p_a}{(p_{N+1}+p_{N+2} ) \ccdot p_a+\xi  p_{N+1} \ccdot p_{N+2}}p_{N+1, \mu} \,p_{N+2, \nu} J^{\mu \nu}_a, \\
S^{(2)} & = \frac{\xi^3}{2} \sum_{a=1}^{N} \frac{1}{(p_{N+1}+p_{N+2} ) \ccdot p_a+\xi  p_{N+1} \ccdot p_{N+2}} \left( \left(p_{N+1, \mu} \,p_{N+2, \nu} J^{\mu \nu}_a\right)^2+\left(\frac{3}{2}-2\kappa\right) ( p_{N+1} \ccdot p_{N+2})^2\right).
\end{align}
Each soft factor $S^{(j)}$ has an expansion in small $\xi$ starting at $\mathcal{O}(\xi^{j+1})$.
The operator $J_a^{\mu\nu}$ appearing in these formulas is the 
spin-0 angular momentum operator,
\be
J^{\mu \nu}_a \equiv p_a^{\mu} \frac{\partial}{\partial p^{\nu}_a}-p_a^{\nu} \frac{\partial}{\partial p^{\mu}_a}.
\ee

\subsubsection*{Double soft theorems with matter}
Since we are able to couple soft scalars to matter fields, we can check whether the double soft theorems are satisfied in this more general case. We saw in the proof of the equivalence principles that there is a universal leading term in the double soft expansion given by Eq.~\eqref{eq:soft-scalar-factorization}, which is indeed equivalent to the leading part of Eq.~\eqref{eq:doublesoft}.
As anticipated, this is the special galileon analogue of Weinberg's universal pole formula \cite{Weinberg:1964ew, Weinberg:1965nx}.

By explicit checks of many examples, we find that also the subleading $\mathcal{O}(\xi^{2\kappa +2})$ part of Eq.~\eqref{eq:doublesoft} continues to hold in the presence of matter, where the angular momentum operator for spinning particles is given by
\be
J^{\mu \nu}_a \equiv p_a^{\mu} \frac{\partial}{\partial p^{\nu}_a}-p_a^{\nu} \frac{\partial}{\partial p^{\mu}_a}+\epsilon_a^{\mu} \frac{\partial}{\partial \epsilon^{\nu}_a}-\epsilon_a^{\nu} \frac{\partial}{\partial \epsilon^{\mu}_a}.
\ee
We have checked this up to eight points for amplitudes involving soft scalars and matter fields with spin up to two. At sub-subleading order, i.e.  $\mathcal{O}(\xi^{2\kappa +3})$, we find that the double soft theorem is no longer universal and depends on the matter fields and their interactions. 
This is consistent with the expectation that only the leading and subleading double soft terms are universal~\cite{Cachazo:2015ksa}, as for the single soft terms in gravity~\cite{Cachazo:2014fwa, Laddha:2017ygw}.

For gravity, in addition to the universal leading soft interaction \eqref{eq:gr-min}, there is a universal subleading soft interaction, 
\begin{align} \label{eq:gr-submin}
\mathcal{V}\big(1_h, \, 2_{\phi^{(s)}}, \, 3_{\phi^{(s)}} \big)&= \frac{2i}{M_{d}^{\frac{D-2}{2}}} ( \epsilon_1 \cdot p_2)^2 (\epsilon_2 \cdot \epsilon_3)^s \nn \\
& +\frac{2si}{M_{d}^{\frac{D-2}{2}}} ( \epsilon_1 \cdot p_2) \left( \epsilon_1 \ccdot \epsilon_2 \, \epsilon_3 \ccdot p_1 +\epsilon_1 \ccdot \epsilon_3 \, \epsilon_2 \ccdot p_3 \right) (\epsilon_2 \cdot \epsilon_3)^{s-1}+\mathcal{O}(p_1^2)\,.
\end{align}
This can be understood from the Lagrangian point of view by noting that nonuniversal terms come from nonminimal interactions involving the Ricci curvature, which are quadratic in the graviton momentum.
By inspection of several examples, we can see that there should also be a universal subleading double soft interaction for the DBI scalar and special galileon coupled to matter,
\begin{align} \label{eq:scalar-submin}
\mathcal{V} \left( 1_\phi, \, 2_\phi,\, 3_{\phi^{(s)}}, \,4_{\phi^{(s)}}\right) =& -\frac{2i\alpha}{\Lambda^{D+2\kappa}} \,  p_{12}^{\kappa} \, p_{13} \, p_{23} \,(\epsilon_3 \cdot \epsilon_4)^s  \\\nonumber
& -\frac{ 2si \alpha}{\Lambda^{D+2\kappa}} \,  p_{12}^{\kappa} \,\left( p_{13} \, \epsilon_3 \ccdot p_2 \, \epsilon_4 \ccdot p_1+ p_{23} \, \epsilon_3 \ccdot p_1 \, \epsilon_4 \ccdot p_2\right)(\epsilon_3 \cdot \epsilon_4)^{s-1}+ \mathcal{O}(\xi^{2\kappa+4})\,,
\end{align}
where $\xi$ counts the total combined power of $p_1$ and $p_2$. We emphasize that although this subleading interaction is present in all of the examples we considered, we have not proven that it, or the subleading double soft theorem, is universal. 

\section{Soft recursion} 
\label{sec:recursion}
In this section we explore yet another interesting similarity between soft scalars and gravity.  Scattering amplitudes in Einstein gravity famously satisfy recursion relations, which can be used to build the higher-point $S$-matrix from knowledge of on-shell processes at lower points. The most well-known of these relations are the celebrated BCFW recursion relations~\cite{Britto:2004ap,Britto:2005fq}. These recursive constructions can, in a sense, be thought of as a direct definition of the $S$-matrix of the theory, without recourse to some underlying Lagrangian description, at least at tree level.

It has recently been understood that scalar field theories that vanish sufficiently quickly in the single soft limit can also obey recursion relations.\footnote{There can also exist recursion relations for theories with nonvanishing soft theorems \cite{Luo:2015tat, Kampf:2019mcd}.} This soft recursion was initially developed in Ref.~\cite{Cheung:2015ota} and was subsequently applied and developed in, e.g., Refs.~\cite{Luo:2015tat, Cheung:2016drk, Padilla:2016mno, Elvang:2018dco, Low:2019ynd}.  It is therefore interesting to understand how the soft behavior of certain DBI or special galileon plus matter amplitudes allows them to be constructed recursively. We begin by briefly reviewing how soft recursion works for the amplitudes of interest in this paper.

\subsection{Review of soft recursion relations}
\label{sec:softbootreview}

Consider an $N$-point amplitude $\mathcal{A}_N$ where the $a$\textsuperscript{th} particle has spin $s_a$, momentum $p_a$, a soft exponent $\sigma_a$,\footnote{We define $\sigma_a$ for massless spin-$s_a$ particles so that the amplitude scales as $\mathcal{O}(p^{\sigma_a +s_a})$ in the soft limit.} and a symmetric traceless polarization tensor $\epsilon^{\mu_1 \dots \mu_{s_a}}_a$. Recall that for each polarization tensor we make the following replacement in the amplitude without any loss of generality:
\be
\epsilon^{\mu_1 \dots \mu_{s_a} }_a \mapsto \epsilon_a^{\mu_1} \dots \epsilon_a^{\mu_{s_a}},
\ee
where $\epsilon^{\mu}_a$ is a null vector.
We now perform a complex deformation of the momenta that rescales the first $N-r$ momenta and shifts the rest,
\begin{subequations} \label{eq:shifts}
\begin{align}
p_a & \mapsto p_a(1-c_a z), &  a&=1, \dots, N-r, \label{eq:ashift} \\
p_a & \mapsto p_a + z q_a, & a&= N-r+1, \dots, N,
\end{align}
\end{subequations}
where $z$ is a complex deformation parameter, $c_a$ are constants, and $q_a$ are constant $D$-vectors.
This is referred to as an ``all-but-$r$-line soft shift."
We take the first $N-r$ particles to be massless, so the on-shell conditions impose the constraints
\begin{subequations}
\label{eq:shiftcons}
\begin{align}
\sum_{a=1}^{N-r} c_a p_a & = \sum_{a=N-r+1}^N q_a, \\
q_a \cdot q_a = q_a \cdot p_a  = q_a \cdot \epsilon_a &=0, \quad{\rm for}\quad a = N-r+1, \dots, N.
\end{align}
\end{subequations}
In total there are $N- r+Dr$ shift variables $c_a$ and $q_a$ subject to $D+3r-n_{r,0}$ constraints, where $n_{r,0}$ is the number of spin-0 particles in the last $r$ legs .
To nontrivially probe the soft kinematics, we require a solution to the constraints \eqref{eq:shiftcons} that is not just an overall rescaling or shift of the momenta; this removes two one-parameter families of solutions, so overall we need
\be \label{eq:rconstraint}
N-r+Dr -2 \geq D+3r-n_{r,0}
\ee
in order to have enough freedom to construct a nontrivial momentum shift.

After shifting the momenta, the amplitude becomes a function of $z$, $\mathcal{A}_N(z)$. 
Using Cauchy's theorem we can write the original amplitude as the contour integral
\be \label{eq:cauchy}
\mathcal{A}_N(0) = \oint_{\gamma} \frac{\mathcal{A}_N(z)}{z F(z) },
\ee
where $\gamma$ is a small contour encircling the origin and $F(z)$ is defined as
\be
F(z) = \prod_{a=1}^{N-r} \left(1-c_a z \right)^{\sigma_a+s_a}.
\label{eq:Ffactor}
\ee 
This denominator of the integrand is chosen such that any would-be poles at $z = c_a^{-1}$ are exactly cancelled by zeros of the numerator because $z \rightarrow c_a^{-1}$ corresponds to a soft limit of the amplitude, and we have assumed that the amplitude has the requisite soft behavior to cancel these poles. 

It is then possible to write the amplitude as minus the sum over the residues of the other singularities of the integrand, assuming that there is no boundary term at $z = \infty$. These singularities correspond to factorization channels of the amplitude, so we can write
\be
\mathcal{A}_N(0) = - \sum_{\substack{{\rm channels} \, I,\\{\rm particles} \, \psi_I}} \res\limits_{z=z_{I\pm}} \left[ \frac{\mathcal{A}_L(z) {\mathcal{A}}_R(z)}{z \left(P_I(z)^2 +m^2_{\psi_I} \right) F(z) } \right],
\label{eq:factorizedamp}
\ee
where the sum runs over all possible factorization channels $I$ where a particle goes on-shell and all possible particles $\psi_I$ that can be exchanged in this channel. 
The factorized amplitudes $\mathcal{A}_L(z)$ and ${\mathcal{A}}_R(z)$ are lower-point amplitudes into which $A_N(z)$ factorizes on a particular channel,  and $P_I(z)$ is the deformation of the sum of momenta $P_I = \sum_{a \in I} p_a$ that add up to zero on the factorization channel. Finally,  $z_{I \pm}$ are the two roots of $P_I(z)^2+m^2_{\psi_I}=0$.

Evaluating the residues leads to the following recursive expression for the amplitude.
\be \label{eq:recursion}
\mathcal{A}_N(0) =\sum_{{\substack{{\rm channels} \, I,\\{\rm particles} \, \psi_I}}} \frac{\mathcal{A}_L(z_{I+}) {\mathcal{A}}_R(z_{I+})}{\left(P_I^2 +m_{\psi_I}^2 \right) (1-z_{I_+}/z_{I_-}) F(z_{I+}) } + \left( z_{I_+} \leftrightarrow z_{I_-}\right).
\ee
In order for the 
recursion relation~\eqref{eq:recursion} to be valid we have to ensure that the integrand of~\eqref{eq:cauchy} goes to zero sufficiently fast as $|z| \rightarrow \infty$. This will be the case if for large $z$ the factors of $z$ in $F(z)$ exceed the factors of $z$ coming from the explicit momenta appearing in the amplitude. The general criteria for this to occur for massless amplitudes in four dimensions is given in Ref.~\cite{Elvang:2018dco}. 

\subsection{All-line soft shift for massless matter amplitudes}

Having reviewed the general formalism of soft recursion, we now apply it to some of the soft scalar plus matter theories discussed in this paper.

We start with the case where all fields are massless, so we can use the all-line soft shift given by \eqref{eq:shifts} with $r=0$.
If the matter fields have vanishing soft behavior, which is the case for photons or derivatively coupled scalars, then the recursion relation based on this shift has greater applicability (in certain dimensions) than the $r>0$ shifts, but it has the disadvantage of working only in dimensions below some upper bound. In particular, by Eq.~\eqref{eq:rconstraint} this momentum shift is possible when
\be \label{eq:allline}
N \geq D+2.
\ee

As an example, consider the case of a free massless scalar or a free photon minimally coupled to the DBI scalar or special galileon. 
For an $N$-point amplitude with $N_{\phi}$ DBI legs ($\kappa=0$) or $N_{\phi}$ special galileon legs ($\kappa=1$), we can choose the denominator function~\eqref{eq:Ffactor} as
\be
F(z) = \prod_{a=1}^{N_{\phi}} \left(1-c_a z \right)^{\kappa+2}  \prod_{a=N_{\phi}+1}^N \left(1-c_a z\right),
\ee
which grows like $z^{N+N_{\phi}(\kappa+1)}$ at large $z$. The $N$-point amplitudes for minimally coupled massless free fields grow  like $p^N$ for DBI and $p^{2N-2}$ for the special galileon, so the absence of a boundary term in the integrand of Eq.~\eqref{eq:cauchy} requires that
\begin{align}
&N_{\phi} > 0 \quad  \rm{for \, \, DBI}, \label{eq:masslessbound-DBI}\\
&N_{\phi} \geq \frac{N}{2} \quad \rm{for \, \, the \, \, special \,\, galileon}. \label{eq:masslessbound}
\end{align}
That is, at least half of the external legs must be special galileons for the recursion relation \eqref{eq:recursion} to be valid in the special galileon theory, but we only need a nonzero number of DBI legs for the all-line soft recursion in the DBI theory. For example, in $D= 4$ we can recursively construct the six-point amplitudes with two external photons or two external massless scalars in both the DBI and special galileon theories, but only in DBI can we recursively construct the six-point amplitudes with four external photons or four external massless scalars. We have explicitly checked that the amplitudes so constructed agree with the expressions computed directly in Section~\ref{sec:minimalcoupling}.

We can understand the special galileon bound~\eqref{eq:masslessbound} from the Lagrangian perspective by considering non-minimal interactions, which are schematically of the form
\be
\Delta \mathcal{L}_{\chi} \sim \nabla^{2k} R^{N_{\phi}/2} (\partial \chi)^{N-N_{\phi}}, \qquad \Delta \mathcal{L}_A \sim \nabla^{2k} R^{N_{\phi}/2} F^{N-N_{\phi}}.
\ee
At $N$ points these produce contact amplitudes that have the same $p^{2N-2}$ momentum scaling as the minimal coupling interactions precisely when
\be
N_{\phi} < \frac{N}{2}.
\ee
So in these cases the soft behavior does not uniquely fix the amplitude, which explains why recursion is not possible. We can similary understand the DBI bound \eqref{eq:masslessbound-DBI} by noting that contact terms can never match the DBI amplitudes for $N_{\phi} >0$.

\subsection{All-but-$r$-line soft shift}

Since DBI and the special galileon have exceptional soft behavior, we can also construct amplitudes involving general matter fields using the all-but-$r$-line soft shift~\cite{Cheung:2016drk, Padilla:2016mno} discussed in Section~\ref{sec:softbootreview}. The recursion relations resulting from this shift are 
valid when the inequality \eqref{eq:rconstraint} is satisfied. This momentum shift is especially suitable when some of the external particles are massive and it works in all dimensions above some lower bound when $r \geq 2$. 

To see how this works,
consider an $N$-point amplitude where the first $N_{\phi}$ fields are DBI scalars or special galileons and perform an $N_{\phi}$-line soft shift on these external legs, with the denominator function
\be
F(z) =\prod_{a=1}^{N_{\phi}} \left(1-c_a z\right)^{\kappa+2},
\ee
which grows like $z^{N_{\phi}(\kappa+2)}$ at infinity. For minimally coupled free fields, the absence of a boundary term requires that
\begin{align}
&N_{\phi} > \frac{N}{2}\quad  \rm{for \ DBI}, \\
&N_{\phi} >\frac{2N-2}{3} \quad \rm{for \ the \ special \ galileon}.
\end{align}
For example, for $D\geq4$ we can recursively construct the six-point amplitudes with two massive matter fields and four DBI or special galileon legs. 
We have explicitly checked this in $D=4$ with matter fields of spin up to two. Note that this recursion also works for general massive higher-spin fields coupled to the DBI scalar or special galileon.

This momentum shift also allows us to recursively build certain amplitudes from minimally coupled matter with self interactions. For example, consider the $(n+2)$-point amplitude with two special galileons and $n$ scalars $\chi$ in the $\chi^n$ scalar theory  considered in Sec.~\ref{sec:chi-soft}. By Eq.~\eqref{eq:rconstraint}, the all-but-$n$-line soft shift is valid for
\be
D \geq \frac{2n}{n-1}. 
\ee
When $n$ is even, this amplitude is not constructible, due to the contributions from interactions connected to the kinetic term. However, for odd $n$ the interactions connected to the kinetic term do not contribute and the amplitude grows like $p^4$, so there is no boundary term in \eqref{eq:cauchy} when we take
\be
F(z) = \left( 1-c_1z \right)^3  \left( 1-c_2 z \right)^3.
\ee
We can thus use recursion to construct these amplitudes for odd $n$ in three or more dimensions, which we have explicitly verified for several values of $n$ and $D$.

\section{``Gravitational" phenomenology}
\label{eq:fakegravity}
%
%
%
Given that soft scalars share so many features with gravity, it is amusing to ponder what a world with a DBI or special galileon as the graviton would be like.\footnote{Needless to say, we are {\it not} advocating that this is how gravity actually behaves.} In this section we indulge this curiosity by deriving the effective gravitational force felt between objects and by exploring some cosmological aspects of the theories.

\subsection{Effective gravitational force}
The fact that DBI and the special galileon couple universally to matter suggests that they should mediate a universal long-range force between matter sources. The long-range $\sim r^{-1}$ potentials mediated by Einstein gravity arise from tree level diagrams with cubic couplings between one massless graviton and two matter particles.  However both DBI and the special galileon are ${\mathbb Z}_2$ invariant, so there are no three-point couplings between these scalars and two matter particles. The long-range potentials therefore arise at one loop, from a diagram of the type depicted in Figure~\ref{fig:forceloop}. We will restrict to deriving the potential between scalar sources, for simplicity.\footnote{Not much generality is lost in the assumption, as the Newtonian potential is not sensitive at leading order to the internal structure of the sources.}

Let ${\cal A}(s,t)$ be the amplitude for elastic scattering of two scalars, of mass $m_1$ and $m_2$.  Let ${\vec{p}}_i$, ${\vec{p}}_f$ be the initial and final spatial momentum of particle 1 in the center of mass frame, and ${\vec q}={\vec p}_f-{\vec p}_i$ the momentum transfer.  We have ${\vec{p}}_i^{\, 2}={\vec{p}}_f^{\, 2}\equiv {\vec p}^{\, 2}$, and the Mandelstam variables can be written in terms of ${\vec{p}}^{\, 2}$ and ${\vec{q}}^{\, 2}$: $t=-{\vec{q}}^{\, 2}$, $s=(m_1+m_2)^2+{\cal O}({\vec{p}}^{\, 2})$.  Let
\be {\cal A}(\vec{q}\, )\equiv\lim_{\vec{p}^{2}\rightarrow 0}{\cal A}(s,t)={\cal A}((m_1+m_2)^2 ,-\vec{q}^{\, 2}).\label{qamplitduedefe}\ee
This is the low-energy limit of the amplitude at fixed momentum transfer.  The static interaction potential is then given by the Fourier transform
\be 
V(r)=-\frac{1}{4m_1m_2}\int \frac{\rd^3 q}{(2\pi)^3}e^{-i\vec{q}\cdot \vec{r}}\mathcal{A}(\vec{q}\,)\,.
\label{eq:potentialFtransform}
\ee

Terms in the amplitude which are analytic in $\vec{q}^{\, 2}$, i.e. analytic in $t$, Fourier transform into delta functions and derivatives of delta functions and so do not contribute to the long-range potential.  Therefore only the parts of the amplitude that are  non-analytic in $t$ are of interest in computing the potential.

\subsubsection*{Gravity}
It is useful to quickly review how the potential in Einstein gravity arises from this on-shell perspective. The long-range gravitational potential between two sources comes from the non-analytic part of the four-point scattering amplitude, which is dominated by the tree-level exchange of virtual gravitons.

Scalar sources interact with gravity through the standard minimal coupling interactions
\be \label{eq:freescalar}
S_{\chi} =- \frac{1}{2}\int\rd^4x\sqrt{-g} \left( g^{\mu \nu} \partial_{\mu} \chi \partial_{\nu} \chi + m^2_{\chi} \chi^2 \right)\,.
\ee
The tree amplitude for scattering two scalars with masses $m_1$ and $m_2$ has only a $t$-channel diagram and is given by
\vspace{-.75cm}
  \be
 \label{eq:tchanpic}
 \begin{tikzpicture}[line width=1.7 pt,baseline={([yshift=-3ex]current bounding box.center)},vertex/.style={anchor=base,
    circle,fill=black!25,minimum size=18pt,inner sep=2pt}]
\draw[dotted] (-1.1,0) -- (1.1,0);
\draw[style] (-1.1,1) -- (1.1,1);
\draw[style={decorate,decoration=complete sines},line width=1] (-.04,0) -- (-.04,2);
\draw[style={decorate,decoration=complete sines},line width=1] (.04,0) -- (.04,2);
\node[scale=1] at (6, .475) {$\displaystyle{=-\frac{1}{M_{\rm Pl}^2 t}\Big(s(s+t)-(m_{1}^2+m_{2}^2)(2s+t)+m_{1}^4+m_{2}^4\Big) }.$};
\end{tikzpicture}
\ee
Up to analytic terms in ${\vec q}^{\, 2}$, the amplitude \eqref{qamplitduedefe} is given by
\be
\mathcal{A}(\vec{q}\,)=\frac{2 \,m_1^2 m_2^2}{M_p^2 \,\vec{q}^{\, 2}},
\ee
and the Fourier transform \eqref{eq:potentialFtransform} to obtain the potential yields the familiar expression for the Newtonian potential between two massive objects,
\be
V(r)=-\frac{m_1 m_2}{8\pi M_{\rm Pl}^2 r}=-\frac{G m_1 m_2}{r}.
\ee

\subsubsection*{DBI}
\begin{figure}[!t]
\begin{center}
\begin{tikzpicture}[line width=1.7 pt,baseline={([yshift=-1.2ex]current bounding box.center)},vertex/.style={anchor=base,
    circle,fill=black!25,minimum size=18pt,inner sep=2pt}]
\draw[dashed] (0,0) ellipse (.4cm and .7cm);
\draw[dotted] (-1.55,-.72) -- (1.55,-.72);
\draw[style] (-1.55,.72) -- (1.55,.72);
\end{tikzpicture}
\end{center}
\caption{\small Loop diagram leading to a potential between two scalar sources $\chi$ and $\psi$, arising from either DBI or the special galileon.}
\label{fig:forceloop}
\end{figure}
We are now ready to turn to the soft scalar cases of interest. In these cases there is no tree-level contribution to the amplitude and the leading long-range force will first arise at the one loop level. We first consider the case of a DBI scalar coupled to matter fields through the minimal coupling as in Eq.~\eqref{eq:dbiscalar}. This is the same as the gravitational minimal coupling but with the metric replaced by the effective DBI metric~\eqref{eq:dbimetric}.
The leading-order contribution to the classical potential arises from the diagram in Figure~\ref{fig:forceloop}.  One-loop forces of this type have also been studied in Refs.~\cite{Kugo:1999mf,Kaloper:2003yf,Brax:2014vva}.

Computing the $t$-channel scattering amplitude at one loop, we obtain
\be
\mathcal{A}^{(t)}\left(1_{\chi}, 2_{\psi}, 3_{\chi}, 4_{\psi} \right)=-\frac{ \alpha^2 t^2 \log(-t)}{3840 \pi^2 \Lambda^{8}}\Big(s^2+t^2+s t+(t-2s)(m_1^2+m_2^2)+m_1^4+m_2^4+4m_1^2 m_2^2 \Big) + \cdots ,
\ee
where we have not shown terms analytic in $t$, which includes the scale of the logarithm and the UV divergences from the loop, since these do not do give rise to long-range forces.  Up to analytic terms in ${\vec q}^{\, 2}$, the amplitude \eqref{qamplitduedefe} is given by
\be \mathcal{A}(\vec{q}\,)=-\frac{\alpha ^2 m_1^2
   m_2^2 }{640 \pi ^2 \Lambda ^8}\vec{q}^{\, 4}\log\left(\vec{q}^{\, 2}\right) +\frac{\alpha ^2 \left(m_1^2+m_1 m_2+m_2^2\right) }{1920 \pi ^2 \Lambda ^8}\vec{q}^{\, 6}\log\left(\vec{q}^{\, 2}\right)-\frac{\alpha ^2 }{3840 \pi ^2 \Lambda ^8}\vec{q}^{\, 8}\log\left(\vec{q}^{\, 2}\right)\,.
   \ee
Now we can take the Fourier transform as in Eq.~\eqref{eq:potentialFtransform} to get the potential.
The Fourier integrals can be performed using the method outlined in the Appendix of Ref.~\cite{Freedman:1991tk}, which yields the result
\be V(r)=-\frac{3 \alpha ^2 m_1 m_2}{128 \pi ^3 \Lambda
   ^8 r^7}-\frac{21 \alpha ^2 \left(m_1^2+m_1
   m_2+m_2^2\right)}{64 \pi ^3 \Lambda ^8 m_1 m_2 r^9} -\frac{189 \alpha ^2}{16 \pi ^3 \Lambda ^8 m_1 m_2 r^{11}}\,.\label{dbipotentiale}
   \ee
Notice that this potential decays very rapidly with distance, like $\sim 1/r^7$, and is universally attractive like gravity.  Note also that since the UV divergences and RG scale of the loop do not contribute to the potential, this is a well defined and calculable quantity in the effective field theory, independent of any UV structure or completion.

\subsubsection*{Special galileon}
We can repeat the same calculation for the special galileon, where scalar sources couple as in Eq.~\eqref{eq:sgalscalar}. We again need to compute the $t$-channel scattering amplitude between unequal mass scalars at one loop. The result up to terms analytic in $t$ is
\be
\mathcal{A}^{(t)}\left(1_{\chi}, 2_{\psi}, 3_{\chi}, 4_{\psi} \right)=-\frac{t^4 \alpha^2 \,\log(-t)}{15360 \pi^2 \Lambda^{12}}\Big(s^2+t^2+s t+(t-2s)(m_1^2+m_2^2)+m_1^4+m_2^4+4m_1^2 m_2^2 \Big)+\cdots.
\ee
We can extract the non-relativistic potential felt by the scalars by taking the low-energy limit, so that the amplitude takes the form
\be \mathcal{A}(\vec q \,)=-\frac{\alpha ^2 m_1^2
   m_2^2 }{2560 \pi ^2 \Lambda ^{12}}{\vec q}^{\, 8} \log\left(\vec{q}^{\, 2}\right)+\frac{\alpha ^2 \left(m_1^2+m_1 m_2+m_2^2\right) }{7680 \pi ^2 \Lambda ^{12}}{\vec q}^{\, 10} \log\left(\vec{q}^{\, 2}\right)-\frac{\alpha ^2 }{15360 \pi ^2 \Lambda ^{12}} {\vec q}^{\, 12} \log\left(\vec{q}^{\, 2}\right)\,,
\ee
up to terms analytic in ${\vec q}^{\, 2}$.
Fourier transforming, we get the potential
\be V(r)=-\frac{567 \alpha ^2 m_1 m_2}{32 \pi ^3 \Lambda
   ^{12} r^{11}} -\frac{10395 \alpha ^2 \left(m_1^2+m_1 m_2+m_2^2\right)}{16 \pi ^3 \Lambda ^{12} m_1 m_2 r^{13}}-\frac{405405 \alpha ^2}{8 \pi ^3 \Lambda ^{12} m_1 m_2 r^{15}}\,.
\ee
This potential falls off with distance very quickly, like $\sim 1/r^{11}$, even faster than the DBI potential in Eq.~\eqref{dbipotentiale}, so gravity in a special galileon world is very weak.

\subsection{Cosmology}
The coarse features of cosmology in models where a soft scalar plays the role of the graviton are rather interesting.
For example, in both the DBI and the special galileon theories there is no CC problem.\footnote{This is a major difference between the models we consider and some of the previous scalar field analogues for gravity~\cite{Sundrum:2003yt,Coradeschi:2013gda,Agrawal:2016ubh}. A motivation for considering these previous models was to shed light on the CC problem. Models based on the conformal dilaton have a precise analogue of the CC problem, essentially because the potential in the theory is not radiatively stable in the presence of matter couplings.}
The analogue of the CC is a term that contains a tadpole ${\cal L}\sim \phi$. For DBI, this tadpole by itself is the full Lagrangian, since it is invariant under the relevant symmetries. For the special galileon, there are in addition compensating galileon terms of odd order which make the action invariant~\cite{Hinterbichler:2015pqa}.   In both cases, these terms cannot be written directly in terms of invariants of the coset construction and so they are Wess--Zumino terms for the relevant symmetries.\footnote{The DBI tadpole can be interpreted geometrically as the volume enclosed by a brane in higher dimensions \cite{Goon:2011qf,Goon:2011uw}.} They are therefore not renormalized either by self-loops, or by loops of heavy fields, so long as we couple to matter in a way that respects the symmetries~\cite{Goon:2016ihr,Preucil:2019nxt}.

Another interesting feature of the special galileon models is that they display a version of degravitation~\cite{Dvali:2002pe,ArkaniHamed:2002fu,Dvali:2007kt}, albeit a version that is too efficient. The special galileon possesses a solution where the field profile is of the form $\phi\sim x^2$, which leads matter fields coupled to the galileon to experience an effective de Sitter geometry if they couple as $\sim\phi T$. However, in the special galileon theory, matter fields couple to the effective metric~\eqref{eq:sgalmetric}, which remains flat.  Additionally, the galileon itself sees a flat metric; even though a tadpole term is not induced radiatively, it has no effect on the dynamics even if it is present---which is a kind of degravitation.

Despite the fact that these theories are not realistic as models of gravity, perhaps there is some lesson to learn for the study of real gravity. In particular, we have seen that these models do not suffer from a CC problem, and display a version of degravitation. Given that these models share many features with gravity, understanding the precise mechanisms for these features could possibly be helpful for the study of gravity itself.

\section{Conclusions}
Scalar field theories with enhanced soft limits have many interesting properties. 
In this paper we have explored how these theories behave when coupled to matter. 
We have seen that the shift symmetries of DBI theory and the special galileon constrain their interactions with matter in a way that is quite similar to the constraints imposed by diffeomorphism invariance when coupling matter to Einstein gravity. 
In particular, we have shown that there are analogues of the $S$-matrix equivalence principle, whereby all matter couples to the DBI scalar or special galileon through a particular quartic vertex with a universal coupling, which can be proven using purely on-shell arguments. These scalar equivalence principles lead to universal double soft theorems that are analogues of Weinberg's soft theorem and, when combined with analogues of the generalized Weinberg-Witten theorem, forbid interactions with massless higher-spin particles. 
We have also seen that soft recursion relations apply to certain amplitudes involving DBI or special galileon legs plus general external matter fields (including massive higher-spin fields), allowing them to be recursively constructed from lower-point amplitudes. 

There are additional aspects to the analogy between gravitation and soft scalar effective field theories that we have not touched on, such as the existence of a Cachazo--He--Yuan (CHY) representation \cite{Cachazo:2014xea} and the double copy.
Another related connection is the transmutation procedure studied in Refs.~\cite{Cheung:2017ems, Cheung:2017yef, Zhou:2018wvn, Bollmann:2018edb}. In this procedure, special galileon amplitudes are produced by applying certain operators to amplitudes of ``extended gravity." It would be interesting to try derive the matter couplings considered here by transmuting gravitational matter interactions. Such a procedure might also shed light on possible UV completions of the special galileon. While positivity constraints show that the galileons in isolation are marginally inconsistent with the existence of an analytic and Lorentz-invariant UV completion~\cite{Adams:2006sv,Bellazzini:2017fep}, adding new modes can alter this conclusion \cite{Cheung:2016yqr, Bonifacio:2016wcb, deRham:2017imi, deRham:2018qqo}. By analogy with gravity, it may be necessary to include an infinite tower of massive higher-spin states to UV complete the special galileon, and such a theory might be obtained by transmuting string theory amplitudes. 
Another question we did not explore in this paper is whether the enhanced soft behavior of scattering amplitudes survives at loop level. Based on the analogy with gravity, our expectation is that the single soft theorems, the soft equivalence principles, and the leading double soft theorems will continue to hold at loop level unless there are anomalies, which might occur when coupling to chiral matter. 

\vspace{-.2cm}
\paragraph{Acknowledgments:} We would like to thank Diederik Roest, Francesco Sgarlata, and David Stefanyszyn for helpful conversations. KH and JB acknowledge support from DOE grant DE-SC0019143 and Simons Foundation Award Number 658908.   
AJ is supported in part by NASA grant NNX16AB27G. 
RAR is supported by DOE grant DE-SC0011941, Simons Foundation Award Number 555117 and NASA grant NNX16AB27G.
JB, KH, LAJ, and AJ would like to thank Daniel Baumann, the Delta-ITP, and the University of Amsterdam for hospitality while part of this work was completed.

\appendix

\section{Degree of freedom counting}
\label{sec:dof}
Thought of as effective field theories defined around flat space, the DBI-matter and special galileon-matter interactions we have constructed in Section~\ref{sec:coupling} and explored in the rest of the paper certainly propagate the correct degrees of freedom in perturbation theory, by construction.
However, we can additionally ask whether there
are extra ghostly degrees of freedom if we trust the classical theories nonlinearly or whether they continue to propagate just the naive degrees of freedom. 

For DBI it is straightforward to see that the minimally coupled scalar and vector theories have second-order equations of motion and so do not propagate extra degrees of freedom nonlinearly.  For the special galileon this is not the case.  In this appendix we show that in the simplest case of a minimally coupled free massless scalar, the theory has an extra ghost degree of freedom.   We expect that this will be the case for more general matter interactions as well.

Consider the Lagrangian defined by Eq.~\eqref{eq:sgalscalar} with $m_{\chi}=0$ and further truncate to mini superspace where the fields involved depend only on time,
\be
\mathcal{L} =   \frac{1}{2} \dot{\phi}^2 + \frac{1}{2} \frac{\dot{\chi}^2}{ \sqrt{1-\alpha \ddot{\phi}^2/\Lambda^{D+2}}}.
\ee
From the perspective of diagnosing an extra degree of freedom this truncation is acceptable, because if we find extra modes here they will also be present when allowing for generic field configurations involving gradients.
The equations of motion for this system are given by
\begin{align}
 \frac{\rd}{\rd t} &\left[ \frac{\dot{\chi}}{ (1-\alpha \ddot{\phi}^2/\Lambda^{D+2})^{1/2}} \right]  = 0,\\
 \frac{\rd^2}{\rd t^2} &\left[ \frac{\alpha\dot{\chi}^2 \ddot{\phi}/\Lambda^{D+2}}{ (1-\alpha \ddot{\phi}^2/\Lambda^{D+2})^{3/2}} \right] -2\ddot{\phi} = 0. \label{eq:EOM2}
\end{align}
The first of these implies that
\be \label{eq:EOMpsi}
\frac{\dot{\chi}}{ (1-\alpha \ddot{\phi}^2/\Lambda^{D+2})^{1/2}} = c_1,
\ee
where $c_1$ is a constant. Substituting this back into Eq.~\eqref{eq:EOM2} gives a fourth-order equation for $\phi$,
\be
 \frac{\rd^2}{\rd t^2} \left[ \frac{\alpha c_1^2 \ddot{\phi}/\Lambda^{D+2}}{ (1-\alpha \ddot{\phi}^2/\Lambda^{D+2})^{1/2}} \right] -2\ddot{\phi} = 0.
\ee
Since this is a fourth order equation, the solution involves four integration constants.  The solution to this can then be substituted into \eqref{eq:EOMpsi}, which becomes a first order equation for $\chi$, which can then be solved for $\chi$ bringing in one more integration constant.

In total we need six independent constants to determine the dynamics, which means that there are six phase space degrees of freedom. Correspondingly there are three physical degrees of freedom, which is one more than in the linearized theory.\footnote{This is in contrast to the toy examples studied in Ref.~\cite{Gabadadze:2012tr}, which have higher-order equations but are structured such that the number of degrees of freedom is still that of a second-order system.} 
Note that from the effective field theory point of view this ghostly degree of freedom does not represent an irremediable sickness but merely signals the breakdown of the effective theory around the cutoff.

\section{Ruling out other quartic vertices}
\label{app:4ptverts}

In this appendix, we
justify our restriction to quartic vertices of the minimal coupling form when finding the leading contributions to the double soft limits in Sections~\ref{sec:DBIequiv} and \ref{sec:sgalequiv}. The essential reason is that other quartic vertices are either subleading in the double soft limit or---if they contribute at the same order in the double soft limit---cannot be made consistent with the required single soft behavior of higher-point amplitudes.  Showing this explicitly requires a careful examination of the possible quartic vertices that contribute at leading order in the double soft limit.

Consider a general on-shell quartic vertex between two DBI scalars or two special galileon scalars and two other particles with spins $s$ and $s'$. The other particles may be non-identical, but we can assume that they have the same mass, since otherwise there is no pole in the double soft limit. We also assume that $s\leq s'$ without loss of generality. The most general on-shell parity-even quartic vertex is then a sum of terms of the form
\be \label{eq:genvertex}
 p_{12}^{r_{12}} p_{13}^{r_{13}} (\epsilon_3 \ccdot \epsilon_4)^{n_{34}} (\epsilon_3 \ccdot p_1)^{m_{31}}   (\epsilon_3 \ccdot p_2)^{m_{32}} (\epsilon_4 \ccdot p_1)^{m_{41}} (\epsilon_4 \ccdot p_2)^{m_{42}},
\ee
where the exponents are non-negative integers satisfying the linear system of equations
\begin{subequations}
\label{eq:linearcons}
\begin{align}
n_{34} + m_{31} +m_{32} & = s, \\
n_{34}+ m_{41} + m_{42} & =s', \\
m_{31}+m_{41}+r_{12}+r_{13} & \geq \kappa+2, \label{eq:ineq1} \\
m_{32}+m_{42}+r_{12}+r_{13} & \geq \kappa+2. \label{eq:ineq2}
\end{align}
\end{subequations}
The first two conditions ensure that the last two particles have the correct spin and the last two conditions ensure that the vertex has at least $\sigma = \kappa+2$ soft behavior on the scalar legs, where $\kappa=0$ for the DBI scalar and $\kappa=1$ for the special galileon. Since the two scalar legs are identical, the vertex should also be symmetric under interchanging $p_1$ and $p_2$ up to on-shell vanishing terms. Note that there can also exist parity-odd quartic vertices with two scalar legs for $D \leq 5$, but here we will restrict to interactions that exist in every dimension.

With these restrictions, we want the vertices with the minimal possible double soft degree, 
\be
\tilde{\sigma} \equiv m_{31}+m_{41} +m_{32}+m_{42}+ 2 r_{12}+r_{13}.
\ee 
Solving the linear system \eqref{eq:linearcons} and imposing the interchange symmetry condition shows that the lowest possible double soft degree is $\tilde{\sigma}=2\kappa+2$. Moreover, for the DBI scalar there is a unique vertex with the minimal double soft degree $\tilde{\sigma}=2$, namely the vertex \eqref{eq:DBI-leading-quartic} of the minimal coupling form. 

For the special galileon, in addition to the vertex of the minimal coupling form \eqref{eq:sgal-min-gen}, there are four other vertices with $\tilde{\sigma}=4$, which have $s'-s$ ranging from zero to two. These additional vertices can be written as
\begin{align}
 {\cal V}_1 &= p_{13}^2 p_{23}^2  (\epsilon_3 \ccdot \epsilon_4)^s, \\
 {\cal V}_2 &=p_{13} p_{23}  (\epsilon_3 \ccdot \epsilon_4)^{s-1}\left( (\epsilon_3 \ccdot p_2) (\epsilon_4 \ccdot p_1) + (\epsilon_3 \ccdot p_1) (\epsilon_4 \ccdot p_2) \right), \\
 {\cal V}_3 &= \left(p^3_{23} \epsilon_4 \ccdot p_1+p^3_{13} \epsilon_4 \ccdot p_2 \right) (\epsilon_3 \ccdot \epsilon_4)^s, \\
 {\cal V}_4 &=p_{13} p_{23} (\epsilon_4 \ccdot p_1) (\epsilon_4 \ccdot p_2)  (\epsilon_3 \ccdot \epsilon_4)^s,
\end{align}
where ${\cal V}_1$ and ${\cal V}_2$ have $s'=s$, ${\cal V}_3$ has $s' = s+1$, and ${\cal V}_4$ has $s'=s+2$. If these vertices were present then they would contribute at leading order in the double soft limit of $\mathcal{A}_{N+2}$ and hence could spoil the special galileon equivalence principle and the universality of the leading double soft theorem. However, we will show that if these extra vertices were present then they would lead to a violation of the $\sigma=3$ single soft behavior.

Suppose, for example, that the $a$\textsuperscript{th} particle with spin $s_a$ interacts with the special galileon through the first two vertices with real coefficients $\beta_a$ and $\gamma_a$, 
\be \label{eq:vertex1}
\delta \mathcal{V}^{(1)}_a =2 i \beta_a p_{13}^2 \,p_{23}^2 \, (\epsilon_3 \ccdot \epsilon_4)^{s_a} +2  i \gamma_a\, p_{13}\, p_{23} \, (\epsilon_3 \ccdot \epsilon_4)^{s_a-1}\Big( (\epsilon_3 \ccdot p_2) (\epsilon_4 \ccdot p_1) + (\epsilon_3 \ccdot p_1) (\epsilon_4 \ccdot p_2) \Big).
\ee
Taking the double and single soft limits of $\mathcal{A}_{N+2}$ in the same way as in Section~\ref{sec:sgalequiv} leads to  the following extra contributions:
\begin{align} \label{eq:softsoft1}
\lim_{\tau \rightarrow 0} \lim_{\xi \rightarrow 0}  \mathcal{A}^{(a)}_{N+2} \Big|_{\beta_a, \, \gamma_a} & = \tau^2 \xi^3\,\gamma_a\, p_{N+1}\ccdot p_{a}\left(p_{N+1}^{\nu_1}p_{N+2}^{\lambda_1}+p_{N+1}^{\lambda_1} p_{N+2}^{\nu_1}\right) \epsilon^{(a)}_{\lambda_1}{}^{\nu_2 \dots \nu_{s_a}}\Pi_{\nu_1 \dots \nu_{s_a}, \mu_1 \dots \mu_{s_a}}({p}_a)\mathcal{A}^{\mu_1 \dots \mu_{s_a}}_{N, a}(p_a)  \nn \\
& +\tau^2 \xi^3\beta_a\, (p_{N+1} \ccdot p_{a})^2 \,p_{N+2} \ccdot p_{a} \mathcal{A}_{N} + \mathcal{O}(\tau^3, \xi^3)+ \mathcal{O}(\xi^4)\, ,
\end{align}
where $\mathcal{A}^{(a)}_{N+2}$ denotes the contribution to $\mathcal{A}_{N+2}$ from the diagram depicted in Figure~\ref{fig:doublesoft}.
The $\sigma=3$ single soft behavior requires eliminating the extra $\mathcal{O}(\tau^2)$ terms and this is only possible if we set $\beta_a =\gamma_a =0$. 

We could try to avoid setting $\gamma_a=0$ by exploiting the same loophole used by the minimal coupling vertices, i.e., by setting $\gamma_a = \gamma$ and summing over $a$. However, this cannot work in this case since the terms multiplying $\gamma_a$ are generically completely different functions of the kinematic variables for each $a$---in particular, if $s_a=0$ these terms vanish since the vertex does not exist for scalars. 
Contact terms also cannot help for the same reason as above, namely that they only contribute even powers of $\xi$.
Another way we could try to avoid this conclusion is by having the $a$\textsuperscript{th} particle interact through vertices of the same form with the special galileon and another particle of the same mass and spin that couples identically to everything else up to signs, since then there would be two similar sets of contributions to Eq.~\eqref{eq:softsoft1} that could be made to cancel one another. However, for real couplings this could only work if the internal particle has a ghostly kinetic term, so this loophole does not work in a unitary theory.

Similar arguments can be used to rule out the two spin-changing vertices, $\mathcal{V}_3$ and $\mathcal{V}_4$. Suppose, for example, that the $a$\textsuperscript{th} particle has spin $s_a$ and interacts with the special galileon and with particles of spin $s_a+1$ and $s_a+2$ through the last two vertices, 
\begin{align} \label{eq:spinchangevertex}
\delta \mathcal{V}^{(2)}_a =&2 i \beta_a \left(p^3_{23} \,\epsilon_4 \ccdot p_1+p^3_{13}\, \epsilon_4 \ccdot p_2 \right) (\epsilon_3 \ccdot \epsilon_4)^{s_a},  \\
\delta \mathcal{V}^{(3)}_a =& 2 i \gamma_a \,p_{13}\, p_{23} (\epsilon_4 \ccdot p_1) (\epsilon_4 \ccdot p_2)  (\epsilon_3 \ccdot \epsilon_4)^{s_a},
\end{align}
where $\beta_a$ and $\gamma_a$ are again real coupling constants (unrelated to the earlier ones). These vertices contribute to $\mathcal{A}_{N+2}$ through the diagrams in Figure~\ref{fig:doublesoft} by exchanging the particles of higher spin. The extra leading contributions after taking the consecutive soft limits are
\begin{align}
\lim_{\tau \rightarrow 0} \lim_{\xi \rightarrow 0} \mathcal{A}^{(a)}_{N+2}\Big|_{\beta_a, \, \gamma_a}  & = \tau^2\xi^3\gamma_a \,p_{N+1}\ccdot p_{a} \epsilon_{(a)}^{\nu_1 \dots \nu_{s_a}}p_{N+1}^{\nu_{s_a+1}}p_{N+2}^{\nu_{s_a+2}} \Pi_{\nu_1 \dots \nu_{s_a+2}, \mu_1 \dots \mu_{s_a+2}}({p}_a)\tilde{\mathcal{A}}^{\mu_1 \dots \mu_{s_a+2}}_{N, a}(p_a) \nn \\
& \quad+ \tau^2\xi^3\beta_a\, (p_{N+2} \ccdot p_a)^2\, \epsilon_{(a)}^{\nu_1  \dots \nu_{s_a}}p_{N+1}^{\nu_{s_a+1}} \Pi_{\nu_1 \dots \nu_{s_a+1}, \mu_1 \dots \mu_{s_a+1}}({p}_a)\tilde{\mathcal{A}}^{\mu_1 \dots \mu_{s_a+1}}_{N,a}(p_a)  \nn \\
& \quad+ \mathcal{O}(\tau^3, \xi^3)+ \mathcal{O}(\xi^4)\, ,
\end{align}
where $\tilde{\mathcal{A}}^{ \mu_1  \dots \mu_{s_a\pm n}}_{N,a}$ is the amplitude related to $\mathcal{A}^{\mu_1\dots \mu_{s_a}}_{N,a}$ by replacing the $a$\textsuperscript{th} leg with the spin-($s_a\pm n$) particle. If the particles of higher spin are also present as external legs in $\mathcal{A}_N$, then the same vertex gives additional contributions to the amplitude. For example, if the $a'$\textsuperscript{th} external leg is the spin-$(s_a+2)$ particle then we get
\begin{align}
\lim_{\tau \rightarrow 0} \lim_{\xi \rightarrow 0} \mathcal{A}^{(a')}_{N+2}\Big|_{\gamma_a}  & = \tau^2\xi^3\gamma_a\,p_{N+1}\ccdot p_{a'}\, \epsilon_{(a')}^{\nu_1\dots \nu_{s_a+2}}p_{N+1, \nu_{s_a+1}}p_{N+2, \nu_{s_a+2}} \Pi_{\nu_1 \dots \nu_{s_a}, \mu_1 \dots \mu_{s_a}}({p}_{a'})\tilde{\mathcal{A}}^{\mu_1 \dots \mu_{s_a}}_{N, a'}(p_{a'}) \nn \\
&\quad+ \mathcal{O}(\tau^3, \xi^3)+ \mathcal{O}(\xi^4)\, .
\end{align}
In order for these extra contributions to not spoil the $\sigma=3$ soft behavior we must again set $\beta_a=\gamma_a=0$. The trick of setting $\gamma_a = \gamma$ and summing over $a$ again cannot work since the accompanying terms depend differently on the kinematics for different legs. This is the same reason why we cannot have cancellations between the different cubic vertices for generic choices of kinematics. Like before, adding multiple particles with the same mass and spin does not help unless they have ghostly kinetic terms.

Finally, note that for all equivalence principle arguments there is an additional step required to show that the two matter particles are identical, since everything up to this point goes through for distinct particles with the same mass and spin that couple symmetrically through the minimal coupling vertex. The resolution is that if all of these particles have healthy kinetic terms, then we can always diagonalize the interactions using the $SO(n)$ symmetry of their kinetic terms.
This then completes the proofs of the soft scalar equivalence principles.

\renewcommand{\em}{}
\bibliographystyle{utphys}
\addcontentsline{toc}{section}{References}
\bibliography{sgal-matter-arxiv-2}

\providecommand{\href}[2]{#2}\begingroup\raggedright\begin{thebibliography}{10}

\bibitem{Cheung:2014dqa}
C.~Cheung, K.~Kampf, J.~Novotny, and J.~Trnka, ``{Effective Field Theories from
  Soft Limits of Scattering Amplitudes},''
  \href{http://dx.doi.org/10.1103/PhysRevLett.114.221602}{{\em Phys. Rev.
  Lett.} {\bf 114} (2015) no.~22, 221602},
\href{http://arxiv.org/abs/1412.4095}{{\tt arXiv:1412.4095 [hep-th]}}.

\bibitem{Cheung:2015ota}
C.~Cheung, K.~Kampf, J.~Novotny, C.-H. Shen, and J.~Trnka, ``{On-Shell
  Recursion Relations for Effective Field Theories},''
  \href{http://dx.doi.org/10.1103/PhysRevLett.116.041601}{{\em Phys. Rev.
  Lett.} {\bf 116} (2016) no.~4, 041601},
\href{http://arxiv.org/abs/1509.03309}{{\tt arXiv:1509.03309 [hep-th]}}.

\bibitem{Luo:2015tat}
H.~Luo and C.~Wen, ``{Recursion relations from soft theorems},''
  \href{http://dx.doi.org/10.1007/JHEP03(2016)088}{{\em JHEP} {\bf 03} (2016)
  088},
\href{http://arxiv.org/abs/1512.06801}{{\tt arXiv:1512.06801 [hep-th]}}.

\bibitem{Cheung:2016drk}
C.~Cheung, K.~Kampf, J.~Novotny, C.-H. Shen, and J.~Trnka, ``{A Periodic Table
  of Effective Field Theories},''
  \href{http://dx.doi.org/10.1007/JHEP02(2017)020}{{\em JHEP} {\bf 02} (2017)
  020},
\href{http://arxiv.org/abs/1611.03137}{{\tt arXiv:1611.03137 [hep-th]}}.

\bibitem{Padilla:2016mno}
A.~Padilla, D.~Stefanyszyn, and T.~Wilson, ``{Probing Scalar Effective Field
  Theories with the Soft Limits of Scattering Amplitudes},''
  \href{http://dx.doi.org/10.1007/JHEP04(2017)015}{{\em JHEP} {\bf 04} (2017)
  015},
\href{http://arxiv.org/abs/1612.04283}{{\tt arXiv:1612.04283 [hep-th]}}.

\bibitem{Elvang:2018dco}
H.~Elvang, M.~Hadjiantonis, C.~R.~T. Jones, and S.~Paranjape, ``{Soft Bootstrap
  and Supersymmetry},'' \href{http://dx.doi.org/10.1007/JHEP01(2019)195}{{\em
  JHEP} {\bf 01} (2019)  195},
\href{http://arxiv.org/abs/1806.06079}{{\tt arXiv:1806.06079 [hep-th]}}.

\bibitem{Low:2019ynd}
I.~Low and Z.~Yin, ``{Soft Bootstrap and Effective Field Theories},''
  \href{http://dx.doi.org/10.1007/JHEP11(2019)078}{{\em JHEP} {\bf 11} (2019)
  078},
\href{http://arxiv.org/abs/1904.12859}{{\tt arXiv:1904.12859 [hep-th]}}.

\bibitem{Rodina:2018pcb}
L.~Rodina, ``{Scattering Amplitudes from Soft Theorems and Infrared
  Behavior},'' \href{http://dx.doi.org/10.1103/PhysRevLett.122.071601}{{\em
  Phys. Rev. Lett.} {\bf 122} (2019) no.~7, 071601},
\href{http://arxiv.org/abs/1807.09738}{{\tt arXiv:1807.09738 [hep-th]}}.

\bibitem{Nicolis:2008in}
A.~Nicolis, R.~Rattazzi, and E.~Trincherini, ``{The Galileon as a local
  modification of gravity},''
  \href{http://dx.doi.org/10.1103/PhysRevD.79.064036}{{\em Phys. Rev.} {\bf
  D79} (2009)  064036},
\href{http://arxiv.org/abs/0811.2197}{{\tt arXiv:0811.2197 [hep-th]}}.

\bibitem{deRham:2010eu}
C.~de~Rham and A.~J. Tolley, ``{DBI and the Galileon reunited},''
  \href{http://dx.doi.org/10.1088/1475-7516/2010/05/015}{{\em JCAP} {\bf 1005}
  (2010)  015},
\href{http://arxiv.org/abs/1003.5917}{{\tt arXiv:1003.5917 [hep-th]}}.

\bibitem{Hinterbichler:2014cwa}
K.~Hinterbichler and A.~Joyce, ``{Goldstones with Extended Shift Symmetries},''
  \href{http://dx.doi.org/10.1142/S0218271814430019}{{\em Int. J. Mod. Phys.}
  {\bf D23} (2014) no.~13, 1443001},
\href{http://arxiv.org/abs/1404.4047}{{\tt arXiv:1404.4047 [hep-th]}}.

\bibitem{Griffin:2014bta}
T.~Griffin, K.~T. Grosvenor, P.~Horava, and Z.~Yan, ``{Scalar Field Theories
  with Polynomial Shift Symmetries},''
  \href{http://dx.doi.org/10.1007/s00220-015-2461-2}{{\em Commun. Math. Phys.}
  {\bf 340} (2015) no.~3, 985--1048},
\href{http://arxiv.org/abs/1412.1046}{{\tt arXiv:1412.1046 [hep-th]}}.

\bibitem{Hinterbichler:2015pqa}
K.~Hinterbichler and A.~Joyce, ``{Hidden symmetry of the Galileon},''
  \href{http://dx.doi.org/10.1103/PhysRevD.92.023503}{{\em Phys. Rev.} {\bf
  D92} (2015) no.~2, 023503},
\href{http://arxiv.org/abs/1501.07600}{{\tt arXiv:1501.07600 [hep-th]}}.

\bibitem{Novotny:2016jkh}
J.~Novotny, ``{Geometry of special Galileons},''
  \href{http://dx.doi.org/10.1103/PhysRevD.95.065019}{{\em Phys. Rev.} {\bf
  D95} (2017) no.~6, 065019},
\href{http://arxiv.org/abs/1612.01738}{{\tt arXiv:1612.01738 [hep-th]}}.

\bibitem{Bogers:2018zeg}
M.~P. Bogers and T.~Brauner, ``{Lie-algebraic classification of effective
  theories with enhanced soft limits},''
  \href{http://dx.doi.org/10.1007/JHEP05(2018)076}{{\em JHEP} {\bf 05} (2018)
  076},
\href{http://arxiv.org/abs/1803.05359}{{\tt arXiv:1803.05359 [hep-th]}}.

\bibitem{Roest:2019oiw}
D.~Roest, D.~Stefanyszyn, and P.~Werkman, ``{An Algebraic Classification of
  Exceptional EFTs},'' \href{http://dx.doi.org/10.1007/JHEP08(2019)081}{{\em
  JHEP} {\bf 08} (2019)  081},
\href{http://arxiv.org/abs/1903.08222}{{\tt arXiv:1903.08222 [hep-th]}}.

\bibitem{Weinberg:1964ew}
S.~Weinberg, ``{Photons and Gravitons in $S$-Matrix Theory: Derivation of
  Charge Conservation and Equality of Gravitational and Inertial Mass},''
\href{http://dx.doi.org/10.1103/PhysRev.135.B1049}{{\em Phys. Rev.} {\bf 135}
  (1964)  B1049--B1056}.

\bibitem{Cachazo:2014fwa}
F.~Cachazo and A.~Strominger, ``{Evidence for a New Soft Graviton Theorem},''
\href{http://arxiv.org/abs/1404.4091}{{\tt arXiv:1404.4091 [hep-th]}}.

\bibitem{Schwab:2014xua}
B.~U.~W. Schwab and A.~Volovich, ``{Subleading Soft Theorem in Arbitrary
  Dimensions from Scattering Equations},''
  \href{http://dx.doi.org/10.1103/PhysRevLett.113.101601}{{\em Phys. Rev.
  Lett.} {\bf 113} (2014) no.~10, 101601},
\href{http://arxiv.org/abs/1404.7749}{{\tt arXiv:1404.7749 [hep-th]}}.

\bibitem{Campiglia:2014yka}
M.~Campiglia and A.~Laddha, ``{Asymptotic symmetries and subleading soft
  graviton theorem},'' \href{http://dx.doi.org/10.1103/PhysRevD.90.124028}{{\em
  Phys. Rev.} {\bf D90} (2014) no.~12, 124028},
\href{http://arxiv.org/abs/1408.2228}{{\tt arXiv:1408.2228 [hep-th]}}.

\bibitem{Britto:2004ap}
R.~Britto, F.~Cachazo, and B.~Feng, ``{New recursion relations for tree
  amplitudes of gluons},''
  \href{http://dx.doi.org/10.1016/j.nuclphysb.2005.02.030}{{\em Nucl. Phys.}
  {\bf B715} (2005)  499--522},
\href{http://arxiv.org/abs/hep-th/0412308}{{\tt arXiv:hep-th/0412308
  [hep-th]}}.

\bibitem{Britto:2005fq}
R.~Britto, F.~Cachazo, B.~Feng, and E.~Witten, ``{Direct proof of tree-level
  recursion relation in Yang-Mills theory},''
  \href{http://dx.doi.org/10.1103/PhysRevLett.94.181602}{{\em Phys. Rev. Lett.}
  {\bf 94} (2005)  181602},
\href{http://arxiv.org/abs/hep-th/0501052}{{\tt arXiv:hep-th/0501052
  [hep-th]}}.

\bibitem{Cachazo:2005ca}
F.~Cachazo and P.~Svrcek, ``{Tree level recursion relations in general
  relativity},''
\href{http://arxiv.org/abs/hep-th/0502160}{{\tt arXiv:hep-th/0502160
  [hep-th]}}.

\bibitem{Benincasa:2007qj}
P.~Benincasa, C.~Boucher-Veronneau, and F.~Cachazo, ``{Taming Tree Amplitudes
  In General Relativity},''
  \href{http://dx.doi.org/10.1088/1126-6708/2007/11/057}{{\em JHEP} {\bf 11}
  (2007)  057},
\href{http://arxiv.org/abs/hep-th/0702032}{{\tt arXiv:hep-th/0702032
  [HEP-TH]}}.

\bibitem{ArkaniHamed:2008yf}
N.~Arkani-Hamed and J.~Kaplan, ``{On Tree Amplitudes in Gauge Theory and
  Gravity},'' \href{http://dx.doi.org/10.1088/1126-6708/2008/04/076}{{\em JHEP}
  {\bf 04} (2008)  076},
\href{http://arxiv.org/abs/0801.2385}{{\tt arXiv:0801.2385 [hep-th]}}.

\bibitem{Cachazo:2015ksa}
F.~Cachazo, S.~He, and E.~Y. Yuan, ``{New Double Soft Emission Theorems},''
  \href{http://dx.doi.org/10.1103/PhysRevD.92.065030}{{\em Phys. Rev.} {\bf
  D92} (2015) no.~6, 065030},
\href{http://arxiv.org/abs/1503.04816}{{\tt arXiv:1503.04816 [hep-th]}}.

\bibitem{Li:2017fsb}
Z.-z. Li, H.-h. Lin, and S.-q. Zhang, ``{On the Symmetry Foundation of Double
  Soft Theorems},'' \href{http://dx.doi.org/10.1007/JHEP12(2017)032}{{\em JHEP}
  {\bf 12} (2017)  032},
\href{http://arxiv.org/abs/1710.00480}{{\tt arXiv:1710.00480 [hep-th]}}.

\bibitem{Guerrieri:2017ujb}
A.~L. Guerrieri, Y.-t. Huang, Z.~Li, and C.~Wen, ``{On the exactness of soft
  theorems},'' \href{http://dx.doi.org/10.1007/JHEP12(2017)052}{{\em JHEP} {\bf
  12} (2017)  052},
\href{http://arxiv.org/abs/1705.10078}{{\tt arXiv:1705.10078 [hep-th]}}.

\bibitem{Sundrum:2003yt}
R.~Sundrum, ``{Gravity's scalar cousin},''
\href{http://arxiv.org/abs/hep-th/0312212}{{\tt arXiv:hep-th/0312212
  [hep-th]}}.

\bibitem{Coradeschi:2013gda}
F.~Coradeschi, P.~Lodone, D.~Pappadopulo, R.~Rattazzi, and L.~Vitale, ``{A
  naturally light dilaton},''
  \href{http://dx.doi.org/10.1007/JHEP11(2013)057}{{\em JHEP} {\bf 11} (2013)
  057},
\href{http://arxiv.org/abs/1306.4601}{{\tt arXiv:1306.4601 [hep-th]}}.

\bibitem{Cachazo:2014xea}
F.~Cachazo, S.~He, and E.~Y. Yuan, ``{Scattering Equations and Matrices: From
  Einstein To Yang-Mills, DBI and NLSM},''
  \href{http://dx.doi.org/10.1007/JHEP07(2015)149}{{\em JHEP} {\bf 07} (2015)
  149},
\href{http://arxiv.org/abs/1412.3479}{{\tt arXiv:1412.3479 [hep-th]}}.

\bibitem{Klein:2015iud}
R.~Klein, M.~Ozkan, and D.~Roest, ``{Galileons as the Scalar Analogue of
  General Relativity},''
  \href{http://dx.doi.org/10.1103/PhysRevD.93.044053}{{\em Phys. Rev.} {\bf
  D93} (2016) no.~4, 044053},
\href{http://arxiv.org/abs/1510.08864}{{\tt arXiv:1510.08864 [hep-th]}}.

\bibitem{Agrawal:2016ubh}
P.~Agrawal and R.~Sundrum, ``{Small Vacuum Energy from Small Equivalence
  Violation in Scalar Gravity},''
  \href{http://dx.doi.org/10.1007/JHEP05(2017)144}{{\em JHEP} {\bf 05} (2017)
  144},
\href{http://arxiv.org/abs/1611.07021}{{\tt arXiv:1611.07021 [hep-th]}}.

\bibitem{Cheung:2016prv}
C.~Cheung and C.-H. Shen, ``{Symmetry for Flavor-Kinematics Duality from an
  Action},'' \href{http://dx.doi.org/10.1103/PhysRevLett.118.121601}{{\em Phys.
  Rev. Lett.} {\bf 118} (2017) no.~12, 121601},
\href{http://arxiv.org/abs/1612.00868}{{\tt arXiv:1612.00868 [hep-th]}}.

\bibitem{Cheung:2017ems}
C.~Cheung, C.-H. Shen, and C.~Wen, ``{Unifying Relations for Scattering
  Amplitudes},'' \href{http://dx.doi.org/10.1007/JHEP02(2018)095}{{\em JHEP}
  {\bf 02} (2018)  095},
\href{http://arxiv.org/abs/1705.03025}{{\tt arXiv:1705.03025 [hep-th]}}.

\bibitem{Cheung:2017yef}
C.~Cheung, G.~N. Remmen, C.-H. Shen, and C.~Wen, ``{Pions as Gluons in Higher
  Dimensions},'' \href{http://dx.doi.org/10.1007/JHEP04(2018)129}{{\em JHEP}
  {\bf 04} (2018)  129},
\href{http://arxiv.org/abs/1709.04932}{{\tt arXiv:1709.04932 [hep-th]}}.

\bibitem{Zhou:2018wvn}
K.~Zhou and B.~Feng, ``{Note on differential operators, CHY integrands, and
  unifying relations for amplitudes},''
  \href{http://dx.doi.org/10.1007/JHEP09(2018)160}{{\em JHEP} {\bf 09} (2018)
  160},
\href{http://arxiv.org/abs/1808.06835}{{\tt arXiv:1808.06835 [hep-th]}}.

\bibitem{Bollmann:2018edb}
M.~Bollmann and L.~Ferro, ``{Transmuting CHY formulae},''
  \href{http://dx.doi.org/10.1007/JHEP01(2019)180}{{\em JHEP} {\bf 01} (2019)
  180},
\href{http://arxiv.org/abs/1808.07451}{{\tt arXiv:1808.07451 [hep-th]}}.

\bibitem{Born:1934gh}
M.~Born and L.~Infeld, ``{Foundations of the new field theory},''
\href{http://dx.doi.org/10.1098/rspa.1934.0059}{{\em Proc. Roy. Soc. Lond.}
  {\bf A144} (1934) no.~852, 425--451}.

\bibitem{Dirac:1962iy}
P.~A.~M. Dirac, ``{An Extensible model of the electron},''
\href{http://dx.doi.org/10.1098/rspa.1962.0124}{{\em Proc. Roy. Soc. Lond.}
  {\bf A268} (1962)  57--67}.

\bibitem{Goon:2011qf}
G.~Goon, K.~Hinterbichler, and M.~Trodden, ``{Symmetries for Galileons and DBI
  scalars on curved space},''
  \href{http://dx.doi.org/10.1088/1475-7516/2011/07/017}{{\em JCAP} {\bf 1107}
  (2011)  017},
\href{http://arxiv.org/abs/1103.5745}{{\tt arXiv:1103.5745 [hep-th]}}.

\bibitem{Gliozzi:2011hj}
F.~Gliozzi, ``{Dirac-Born-Infeld action from spontaneous breakdown of Lorentz
  symmetry in brane-world scenarios},''
  \href{http://dx.doi.org/10.1103/PhysRevD.84.027702}{{\em Phys. Rev.} {\bf
  D84} (2011)  027702},
\href{http://arxiv.org/abs/1103.5377}{{\tt arXiv:1103.5377 [hep-th]}}.

\bibitem{Goon:2012dy}
G.~Goon, K.~Hinterbichler, A.~Joyce, and M.~Trodden, ``{Galileons as
  Wess-Zumino Terms},'' \href{http://dx.doi.org/10.1007/JHEP06(2012)004}{{\em
  JHEP} {\bf 06} (2012)  004},
\href{http://arxiv.org/abs/1203.3191}{{\tt arXiv:1203.3191 [hep-th]}}.

\bibitem{deRham:2013hsa}
C.~de~Rham, M.~Fasiello, and A.~J. Tolley, ``{Galileon Duality},''
  \href{http://dx.doi.org/10.1016/j.physletb.2014.03.061}{{\em Phys. Lett.}
  {\bf B733} (2014)  46--51},
\href{http://arxiv.org/abs/1308.2702}{{\tt arXiv:1308.2702 [hep-th]}}.

\bibitem{Kampf:2014rka}
K.~Kampf and J.~Novotny, ``{Unification of Galileon Dualities},''
  \href{http://dx.doi.org/10.1007/JHEP10(2014)006}{{\em JHEP} {\bf 10} (2014)
  006},
\href{http://arxiv.org/abs/1403.6813}{{\tt arXiv:1403.6813 [hep-th]}}.

\bibitem{Garcia-Saenz:2019yok}
S.~Garcia-Saenz, J.~Kang, and R.~Penco, ``{Gauged Galileons},''
  \href{http://dx.doi.org/10.1007/JHEP07(2019)081}{{\em JHEP} {\bf 07} (2019)
  081},
\href{http://arxiv.org/abs/1905.05190}{{\tt arXiv:1905.05190 [hep-th]}}.

\bibitem{Carrillo-Gonzalez:2019aao}
M.~Carrillo~González, R.~Penco, and M.~Trodden, ``{Shift symmetries, soft
  limits, and the double copy beyond leading order},''
\href{http://arxiv.org/abs/1908.07531}{{\tt arXiv:1908.07531 [hep-th]}}.

\bibitem{Bernard:2014bfa}
L.~Bernard, C.~Deffayet, and M.~von Strauss, ``{Consistent massive graviton on
  arbitrary backgrounds},''
  \href{http://dx.doi.org/10.1103/PhysRevD.91.104013}{{\em Phys. Rev.} {\bf
  D91} (2015) no.~10, 104013},
\href{http://arxiv.org/abs/1410.8302}{{\tt arXiv:1410.8302 [hep-th]}}.

\bibitem{Bernard:2015mkk}
L.~Bernard, C.~Deffayet, and M.~von Strauss, ``{Massive graviton on arbitrary
  background: derivation, syzygies, applications},''
  \href{http://dx.doi.org/10.1088/1475-7516/2015/06/038}{{\em JCAP} {\bf 1506}
  (2015)  038},
\href{http://arxiv.org/abs/1504.04382}{{\tt arXiv:1504.04382 [hep-th]}}.

\bibitem{Bernard:2015uic}
L.~Bernard, C.~Deffayet, A.~Schmidt-May, and M.~von Strauss, ``{Linear spin-2
  fields in most general backgrounds},''
  \href{http://dx.doi.org/10.1103/PhysRevD.93.084020}{{\em Phys. Rev.} {\bf
  D93} (2016) no.~8, 084020},
\href{http://arxiv.org/abs/1512.03620}{{\tt arXiv:1512.03620 [hep-th]}}.

\bibitem{Bernard:2017tcg}
L.~Bernard, C.~Deffayet, K.~Hinterbichler, and M.~von Strauss, ``{Partially
  Massless Graviton on Beyond Einstein Spacetimes},''
  \href{http://dx.doi.org/10.1103/PhysRevD.98.069902,
  10.1103/PhysRevD.95.124036}{{\em Phys. Rev.} {\bf D95} (2017) no.~12,
  124036}, \href{http://arxiv.org/abs/1703.02538}{{\tt arXiv:1703.02538
  [hep-th]}}.
[Erratum: Phys. Rev.D98,no.6,069902(2018)].

\bibitem{deRham:2010kj}
C.~de~Rham, G.~Gabadadze, and A.~J. Tolley, ``{Resummation of Massive
  Gravity},'' \href{http://dx.doi.org/10.1103/PhysRevLett.106.231101}{{\em
  Phys. Rev. Lett.} {\bf 106} (2011)  231101},
\href{http://arxiv.org/abs/1011.1232}{{\tt arXiv:1011.1232 [hep-th]}}.

\bibitem{Boulanger:2000rq}
N.~Boulanger, T.~Damour, L.~Gualtieri, and M.~Henneaux, ``{Inconsistency of
  interacting, multigraviton theories},''
  \href{http://dx.doi.org/10.1016/S0550-3213(00)00718-5}{{\em Nucl. Phys.} {\bf
  B597} (2001)  127--171},
\href{http://arxiv.org/abs/hep-th/0007220}{{\tt arXiv:hep-th/0007220
  [hep-th]}}.

\bibitem{Hinterbichler:2010xn}
K.~Hinterbichler, M.~Trodden, and D.~Wesley, ``{Multi-field galileons and
  higher co-dimension branes},''
  \href{http://dx.doi.org/10.1103/PhysRevD.82.124018}{{\em Phys. Rev.} {\bf
  D82} (2010)  124018},
\href{http://arxiv.org/abs/1008.1305}{{\tt arXiv:1008.1305 [hep-th]}}.

\bibitem{Cachazo:2016njl}
F.~Cachazo, P.~Cha, and S.~Mizera, ``{Extensions of Theories from Soft
  Limits},'' \href{http://dx.doi.org/10.1007/JHEP06(2016)170}{{\em JHEP} {\bf
  06} (2016)  170},
\href{http://arxiv.org/abs/1604.03893}{{\tt arXiv:1604.03893 [hep-th]}}.

\bibitem{Arkani-Hamed:2016rak}
N.~Arkani-Hamed, L.~Rodina, and J.~Trnka, ``{Locality and Unitarity of
  Scattering Amplitudes from Singularities and Gauge Invariance},''
  \href{http://dx.doi.org/10.1103/PhysRevLett.120.231602}{{\em Phys. Rev.
  Lett.} {\bf 120} (2018) no.~23, 231602},
\href{http://arxiv.org/abs/1612.02797}{{\tt arXiv:1612.02797 [hep-th]}}.

\bibitem{Kampf:2019mcd}
K.~Kampf, J.~Novotny, M.~Shifman, and J.~Trnka, ``{New Soft Theorems for
  Goldstone Boson Amplitudes},''
  \href{http://dx.doi.org/10.1103/PhysRevLett.124.111601}{{\em Phys. Rev.
  Lett.} {\bf 124} (2020) no.~11, 111601},
\href{http://arxiv.org/abs/1910.04766}{{\tt arXiv:1910.04766 [hep-th]}}.

\bibitem{Aragone:1979hx}
C.~Aragone and S.~Deser, ``{Consistency Problems of Hypergravity},''
\href{http://dx.doi.org/10.1016/0370-2693(79)90808-6}{{\em Phys. Lett.} {\bf
  86B} (1979)  161--163}.

\bibitem{Metsaev:2005ar}
R.~R. Metsaev, ``{Cubic interaction vertices of massive and massless higher
  spin fields},'' \href{http://dx.doi.org/10.1016/j.nuclphysb.2006.10.002}{{\em
  Nucl. Phys.} {\bf B759} (2006)  147--201},
\href{http://arxiv.org/abs/hep-th/0512342}{{\tt arXiv:hep-th/0512342
  [hep-th]}}.

\bibitem{Porrati:2008rm}
M.~Porrati, ``{Universal Limits on Massless High-Spin Particles},''
  \href{http://dx.doi.org/10.1103/PhysRevD.78.065016}{{\em Phys. Rev.} {\bf
  D78} (2008)  065016},
\href{http://arxiv.org/abs/0804.4672}{{\tt arXiv:0804.4672 [hep-th]}}.

\bibitem{Boulanger:2008tg}
N.~Boulanger, S.~Leclercq, and P.~Sundell, ``{On The Uniqueness of Minimal
  Coupling in Higher-Spin Gauge Theory},''
  \href{http://dx.doi.org/10.1088/1126-6708/2008/08/056}{{\em JHEP} {\bf 08}
  (2008)  056},
\href{http://arxiv.org/abs/0805.2764}{{\tt arXiv:0805.2764 [hep-th]}}.

\bibitem{Taronna:2011kt}
M.~Taronna, ``{Higher-Spin Interactions: four-point functions and beyond},''
  \href{http://dx.doi.org/10.1007/JHEP04(2012)029}{{\em JHEP} {\bf 04} (2012)
  029},
\href{http://arxiv.org/abs/1107.5843}{{\tt arXiv:1107.5843 [hep-th]}}.

\bibitem{Metsaev:1991mt}
R.~R. Metsaev, ``{Poincare invariant dynamics of massless higher spins: Fourth
  order analysis on mass shell},''
\href{http://dx.doi.org/10.1142/S0217732391000348}{{\em Mod. Phys. Lett.} {\bf
  A6} (1991)  359--367}.

\bibitem{Metsaev:1991nb}
R.~R. Metsaev, ``{S matrix approach to massless higher spins theory. 2: The
  Case of internal symmetry},''
\href{http://dx.doi.org/10.1142/S0217732391002839}{{\em Mod. Phys. Lett.} {\bf
  A6} (1991)  2411--2421}.

\bibitem{Ponomarev:2016lrm}
D.~Ponomarev and E.~D. Skvortsov, ``{Light-Front Higher-Spin Theories in Flat
  Space},'' \href{http://dx.doi.org/10.1088/1751-8121/aa56e7}{{\em J. Phys.}
  {\bf A50} (2017) no.~9, 095401},
\href{http://arxiv.org/abs/1609.04655}{{\tt arXiv:1609.04655 [hep-th]}}.

\bibitem{Skvortsov:2018jea}
E.~D. Skvortsov, T.~Tran, and M.~Tsulaia, ``{Quantum Chiral Higher Spin
  Gravity},'' \href{http://dx.doi.org/10.1103/PhysRevLett.121.031601}{{\em
  Phys. Rev. Lett.} {\bf 121} (2018) no.~3, 031601},
\href{http://arxiv.org/abs/1805.00048}{{\tt arXiv:1805.00048 [hep-th]}}.

\bibitem{Sleight:2016xqq}
C.~Sleight and M.~Taronna, ``{Higher-Spin Algebras, Holography and Flat
  Space},'' \href{http://dx.doi.org/10.1007/JHEP02(2017)095}{{\em JHEP} {\bf
  02} (2017)  095},
\href{http://arxiv.org/abs/1609.00991}{{\tt arXiv:1609.00991 [hep-th]}}.

\bibitem{Taronna:2017wbx}
M.~Taronna, ``{On the Non-Local Obstruction to Interacting Higher Spins in Flat
  Space},'' \href{http://dx.doi.org/10.1007/JHEP05(2017)026}{{\em JHEP} {\bf
  05} (2017)  026},
\href{http://arxiv.org/abs/1701.05772}{{\tt arXiv:1701.05772 [hep-th]}}.

\bibitem{Weinberg:1965nx}
S.~Weinberg, ``{Infrared photons and gravitons},''
\href{http://dx.doi.org/10.1103/PhysRev.140.B516}{{\em Phys. Rev.} {\bf 140}
  (1965)  B516--B524}.

\bibitem{Strominger:2017zoo}
A.~Strominger, ``{Lectures on the Infrared Structure of Gravity and Gauge
  Theory},''
\href{http://arxiv.org/abs/1703.05448}{{\tt arXiv:1703.05448 [hep-th]}}.

\bibitem{Campiglia:2017dpg}
M.~Campiglia, L.~Coito, and S.~Mizera, ``{Can scalars have asymptotic
  symmetries?},'' \href{http://dx.doi.org/10.1103/PhysRevD.97.046002}{{\em
  Phys. Rev.} {\bf D97} (2018) no.~4, 046002},
\href{http://arxiv.org/abs/1703.07885}{{\tt arXiv:1703.07885 [hep-th]}}.

\bibitem{Hamada:2017atr}
Y.~Hamada and S.~Sugishita, ``{Soft pion theorem, asymptotic symmetry and new
  memory effect},'' \href{http://dx.doi.org/10.1007/JHEP11(2017)203}{{\em JHEP}
  {\bf 11} (2017)  203},
\href{http://arxiv.org/abs/1709.05018}{{\tt arXiv:1709.05018 [hep-th]}}.

\bibitem{Campiglia:2017xkp}
M.~Campiglia and L.~Coito, ``{Asymptotic charges from soft scalars in even
  dimensions},'' \href{http://dx.doi.org/10.1103/PhysRevD.97.066009}{{\em Phys.
  Rev.} {\bf D97} (2018) no.~6, 066009},
\href{http://arxiv.org/abs/1711.05773}{{\tt arXiv:1711.05773 [hep-th]}}.

\bibitem{Laddha:2017ygw}
A.~Laddha and A.~Sen, ``{Sub-subleading Soft Graviton Theorem in Generic
  Theories of Quantum Gravity},''
  \href{http://dx.doi.org/10.1007/JHEP10(2017)065}{{\em JHEP} {\bf 10} (2017)
  065},
\href{http://arxiv.org/abs/1706.00759}{{\tt arXiv:1706.00759 [hep-th]}}.

\bibitem{Kugo:1999mf}
T.~Kugo and K.~Yoshioka, ``{Probing extra dimensions using Nambu-Goldstone
  bosons},'' \href{http://dx.doi.org/10.1016/S0550-3213(00)00645-3}{{\em Nucl.
  Phys.} {\bf B594} (2001)  301--328},
\href{http://arxiv.org/abs/hep-ph/9912496}{{\tt arXiv:hep-ph/9912496
  [hep-ph]}}.

\bibitem{Kaloper:2003yf}
N.~Kaloper, ``{Disformal inflation},''
  \href{http://dx.doi.org/10.1016/j.physletb.2004.01.005}{{\em Phys. Lett.}
  {\bf B583} (2004)  1--13},
\href{http://arxiv.org/abs/hep-ph/0312002}{{\tt arXiv:hep-ph/0312002
  [hep-ph]}}.

\bibitem{Brax:2014vva}
P.~Brax and C.~Burrage, ``{Constraining Disformally Coupled Scalar Fields},''
  \href{http://dx.doi.org/10.1103/PhysRevD.90.104009}{{\em Phys. Rev.} {\bf
  D90} (2014) no.~10, 104009},
\href{http://arxiv.org/abs/1407.1861}{{\tt arXiv:1407.1861 [astro-ph.CO]}}.

\bibitem{Freedman:1991tk}
D.~Z. Freedman, K.~Johnson, and J.~I. Latorre, ``{Differential regularization
  and renormalization: A New method of calculation in quantum field theory},''
\href{http://dx.doi.org/10.1016/0550-3213(92)90240-C}{{\em Nucl. Phys.} {\bf
  B371} (1992)  353--414}.

\bibitem{Goon:2011uw}
G.~Goon, K.~Hinterbichler, and M.~Trodden, ``{A New Class of Effective Field
  Theories from Embedded Branes},''
  \href{http://dx.doi.org/10.1103/PhysRevLett.106.231102}{{\em Phys. Rev.
  Lett.} {\bf 106} (2011)  231102},
\href{http://arxiv.org/abs/1103.6029}{{\tt arXiv:1103.6029 [hep-th]}}.

\bibitem{Goon:2016ihr}
G.~Goon, K.~Hinterbichler, A.~Joyce, and M.~Trodden, ``{Aspects of Galileon
  Non-Renormalization},'' \href{http://dx.doi.org/10.1007/JHEP11(2016)100}{{\em
  JHEP} {\bf 11} (2016)  100},
\href{http://arxiv.org/abs/1606.02295}{{\tt arXiv:1606.02295 [hep-th]}}.

\bibitem{Preucil:2019nxt}
F.~Přeučil and J.~Novotný, ``{Special Galileon at one loop},''
  \href{http://dx.doi.org/10.1007/JHEP11(2019)166}{{\em JHEP} {\bf 11} (2019)
  166},
\href{http://arxiv.org/abs/1909.06214}{{\tt arXiv:1909.06214 [hep-th]}}.

\bibitem{Dvali:2002pe}
G.~Dvali, G.~Gabadadze, and M.~Shifman, ``{Diluting cosmological constant in
  infinite volume extra dimensions},''
  \href{http://dx.doi.org/10.1103/PhysRevD.67.044020}{{\em Phys. Rev.} {\bf
  D67} (2003)  044020},
\href{http://arxiv.org/abs/hep-th/0202174}{{\tt arXiv:hep-th/0202174
  [hep-th]}}.

\bibitem{ArkaniHamed:2002fu}
N.~Arkani-Hamed, S.~Dimopoulos, G.~Dvali, and G.~Gabadadze, ``{Nonlocal
  modification of gravity and the cosmological constant problem},''
\href{http://arxiv.org/abs/hep-th/0209227}{{\tt arXiv:hep-th/0209227
  [hep-th]}}.

\bibitem{Dvali:2007kt}
G.~Dvali, S.~Hofmann, and J.~Khoury, ``{Degravitation of the cosmological
  constant and graviton width},''
  \href{http://dx.doi.org/10.1103/PhysRevD.76.084006}{{\em Phys. Rev.} {\bf
  D76} (2007)  084006},
\href{http://arxiv.org/abs/hep-th/0703027}{{\tt arXiv:hep-th/0703027
  [HEP-TH]}}.

\bibitem{Adams:2006sv}
A.~Adams, N.~Arkani-Hamed, S.~Dubovsky, A.~Nicolis, and R.~Rattazzi,
  ``{Causality, analyticity and an IR obstruction to UV completion},''
  \href{http://dx.doi.org/10.1088/1126-6708/2006/10/014}{{\em JHEP} {\bf 10}
  (2006)  014},
\href{http://arxiv.org/abs/hep-th/0602178}{{\tt arXiv:hep-th/0602178
  [hep-th]}}.

\bibitem{Bellazzini:2017fep}
B.~Bellazzini, F.~Riva, J.~Serra, and F.~Sgarlata, ``{Beyond Positivity Bounds
  and the Fate of Massive Gravity},''
  \href{http://dx.doi.org/10.1103/PhysRevLett.120.161101}{{\em Phys. Rev.
  Lett.} {\bf 120} (2018) no.~16, 161101},
\href{http://arxiv.org/abs/1710.02539}{{\tt arXiv:1710.02539 [hep-th]}}.

\bibitem{Cheung:2016yqr}
C.~Cheung and G.~N. Remmen, ``{Positive Signs in Massive Gravity},''
  \href{http://dx.doi.org/10.1007/JHEP04(2016)002}{{\em JHEP} {\bf 04} (2016)
  002},
\href{http://arxiv.org/abs/1601.04068}{{\tt arXiv:1601.04068 [hep-th]}}.

\bibitem{Bonifacio:2016wcb}
J.~Bonifacio, K.~Hinterbichler, and R.~A. Rosen, ``{Positivity constraints for
  pseudolinear massive spin-2 and vector Galileons},''
  \href{http://dx.doi.org/10.1103/PhysRevD.94.104001}{{\em Phys. Rev.} {\bf
  D94} (2016) no.~10, 104001},
\href{http://arxiv.org/abs/1607.06084}{{\tt arXiv:1607.06084 [hep-th]}}.

\bibitem{deRham:2017imi}
C.~de~Rham, S.~Melville, A.~J. Tolley, and S.-Y. Zhou, ``{Massive Galileon
  Positivity Bounds},'' \href{http://dx.doi.org/10.1007/JHEP09(2017)072}{{\em
  JHEP} {\bf 09} (2017)  072},
\href{http://arxiv.org/abs/1702.08577}{{\tt arXiv:1702.08577 [hep-th]}}.

\bibitem{deRham:2018qqo}
C.~de~Rham, S.~Melville, A.~J. Tolley, and S.-Y. Zhou, ``{Positivity Bounds for
  Massive Spin-1 and Spin-2 Fields},''
  \href{http://dx.doi.org/10.1007/JHEP03(2019)182}{{\em JHEP} {\bf 03} (2019)
  182},
\href{http://arxiv.org/abs/1804.10624}{{\tt arXiv:1804.10624 [hep-th]}}.

\bibitem{Gabadadze:2012tr}
G.~Gabadadze, K.~Hinterbichler, J.~Khoury, D.~Pirtskhalava, and M.~Trodden,
  ``{A Covariant Master Theory for Novel Galilean Invariant Models and Massive
  Gravity},'' \href{http://dx.doi.org/10.1103/PhysRevD.86.124004}{{\em Phys.
  Rev.} {\bf D86} (2012)  124004},
\href{http://arxiv.org/abs/1208.5773}{{\tt arXiv:1208.5773 [hep-th]}}.

\end{thebibliography}\endgroup

\end{document}